\documentclass[12pt]{article}
\usepackage{amsfonts}
\usepackage{bbm}
\usepackage{graphicx}
\usepackage{booktabs}
\usepackage{subfigure}
\usepackage{mathrsfs}
\usepackage{color}
\usepackage{verbatim}
\usepackage{cancel}
\usepackage{geometry}             
\usepackage{amssymb}
\usepackage{amsmath}
\usepackage{epstopdf}
\usepackage{cases}
\usepackage{cite}
\usepackage{setspace}
\usepackage{ulem}

\usepackage[hyperindex=true,
          pdfstartview=FitH,
          bookmarksnumbered=true,
          bookmarksopen=true,
          citecolor=blue,
          linkcolor=blue,
          colorlinks=true,
          urlcolor=rossoCP3,
          unicode]{hyperref}

\definecolor{rossoCP3}{cmyk}{0,.88,.77,.40}

\begin{document}

\title{\bf \textsf{Ruppeiner geometry and thermodynamic phase transition of the black hole in massive gravity}}
{\author{\small Bin Wu${}^{1,2,3,4}$\thanks{{\em email}: \href{mailto:binwu@nwu.edu.cn}{binwu@nwu.edu.cn}}{ }, Chao Wang${}^{1}$\thanks{{\em email}: \href{mailto:chaowang1510@gmail.com}{chaowang1510@gmail.com}}{ }, Zhen-Ming Xu${}^{1,2,3,4}$\thanks{{\em email}: \href{mailto:xuzhenm@nwu.edu.cn}{xuzhenm@nwu.edu.cn, corresponding author}}{}, and Wen-Li Yang${}^{1,2,3,4}$\thanks{{\em email}: \href{mailto:wlyang@nwu.edu.cn}{wlyang@nwu.edu.cn}}
\vspace{5pt}\\
\small $^{1}${\it School of Physics, Northwest University, Xi'an 710127, China}\\
\small $^{2}${\it Institute of Modern Physics, Northwest University, Xi'an 710127, China}\\
\small $^{3}${\it Shaanxi Key Laboratory for Theoretical Physics Frontiers, Xi'an 710127, China}\\
\small $^{4}${\it Peng Huanwu Center for Fundamental Theory, Xi'an 710127, China}
}

\date{}
\maketitle
\begin{spacing}{1.2}
\begin{abstract}
The phase transition and thermodynamic geometry of a 4-dimensional AdS topological charged black hole in de Rham, Gabadadze and Tolley (dRGT) massive gravity have been studied. After introducing a normalized thermodynamic scalar curvature, it is speculated that its value is related to the interaction between the underlying black hole molecules if the black hole molecules exist. We show that there does exist a crucial parameter given in terms of the topology, charge, and massive parameters of the black hole, which characterizes the thermodynamic properties of the black hole. It is found that when the parameter is positive, the singlet large black hole phase does not exist for sufficient low temperature and there is a weak repulsive interaction dominating for the small black hole which is similar to the Reissner-Nordstr\"{o}m AdS black hole; when the parameter is negative, an additional phase region describing large black holes also implies a dominant repulsive interaction. These constitute the distinguishable features of dRGT massive topological black hole from those of the Reissner-Nordstr\"{o}m AdS black hole as well as the Van der Waals fluid system.
\end{abstract}

\section{\textsf{Introduction}}
It hints a fascinating set of connections between black hole physics and thermodynamics since the pioneering works of Hawking and Bekenstein about the temperature and entropy of black holes\cite{Bekenstein:1973ur,Hawking:1974sw,Bardeen:1973gs,Hawking:1976de}. The thermodynamics theory, which is widely  used in ordinary systems, is now a useful tool for us to understand black hole systems. Inspired by the Anti-de Sitter/Conformal Field Theory (AdS/CFT) correspondence\cite{Maldacena:1997re,Gubser:1998bc,Witten:1998qj}, one can relate the Hawking-Page phase transition\cite{Hawking:1982dh} with the confinement/deconfinement phase transition of gauge field\cite{Witten:1998zw}. Then particular attention has been paid to the phase transition of black holes in the AdS spacetime. Recently, the introduction of variation of cosmological constant $\Lambda$ in the first law of black hole thermodynamics has been extensively studied in literatures\cite{Kastor:2009wy,Cvetic:2010jb,Dolan:2011xt,Dolan:2013ft,El-Menoufi:2013pza,Castro:2013pqa}. Such a variation seems to be awkward, as we are considering black hole ensembles with different theories, while the standard thermodynamics is concerned with changes of the state of the system in a given theory. However, there are some considerable reasons why the variation of $\Lambda$ is feasible. i) At first, a number of researches support for variable $\Lambda$, for example, the four-dimensional cosmological constant represents the varying energy density of a $4-$form gauge field strength\cite{Brown:1987dd,Brown:1988kg}. ii) What's more, one may consider  `more fundamental' theories, and the general gravity theory we considered could be regarded as the effective field theory of supergravity theory. The cosmological constant could be regarded as a parameter in a solution of 10-dimensional supergravity, and is no more fundamental to the theory than a black hole mass or any other parameter in the metric. The cosmological constant or the other physical constants are not fixed a priori, but arise as vacuum expectation values and hence can vary, so that we are always referring to the same field theory in this sense. iii) In the presence of a cosmological constant, the scaling argument is no longer valid and the Smarr relation would be inconsistent with the thermodynamic first law of black hole if the $\Lambda$ is fixed. Meanwhile, the pressure-volume term, which is necessary to appear in everyday thermodynamics, would be naturally introduced if the negative cosmological constant is associated with the positive pressure as $P=-\frac{\Lambda}{8\pi G}$ and its conjugate quantity is the thermodynamic volume. This is also a practical reason for the Smarr relation to holding in the case of asymptotically AdS black holes. Hence the thermodynamics of the AdS black hole with varying cosmological constant would be reasonable.

The introduction of the extended phase space enriches the behaviors of the black hole thermodynamics, like that the small/large phase transition of the charged AdS black hole is found in analogy with the liquid/gas phase transition of Van der Waals system\cite{Kubiznak:2012wp}. In the line of this approach, many efforts have been done to reveal the abundant phase structure and critical phenomena of black holes in such an extended phase space\cite{Toledo:2019amt,Hendi:2012um,Wei:2012ui,Cai:2013qga,Zhao:2013oza,Mo:2013ela,Altamirano:2013ane,Spallucci:2013osa,Xu:2014tja,Miao:2018fke,Miao:2016ulg,Xu:2013zea}. This intriguing subject is come to know as black hole chemistry, which associates with each black hole parameter a chemical equivalent in representations of the first law of black hole thermodynamics\cite{Frassino:2015oca,Kubiznak:2016qmn}.

Despite the immense conceptual and phenomenological success of black hole thermodynamics, the description of the microstructure of the black hole remains a huge challenge in current researches. In the literature\cite{Wei:2015iwa}, certain unknown black hole micromolecules are proposed, which provides some intuitions and new insights for the underlying microstructure of black holes. In this significant approach, the Ruppeiner geometry and the number density of the microstates of the black hole are introduced. Originally, Ruppeiner geometry is mainly to use the Hessian matrix structure to represent the thermodynamic fluctuation theory\cite{PhysRevA.24.488,Ruppeiner:1983zz,Ruppeiner:1995zz}. A Ruppeiner metric (line element) measuring the distance between two neighboring fluctuation states is defined. The sign of the Ruppeiner scalar curvature may qualitatively imply the interaction between the modules of the fluid system. Specifically, a positive (or negative) thermodynamic scalar curvature maybe indicate a repulsive (or attractive) interaction. Moreover, a noninteracting system such as the ideal gas corresponds to the flat Ruppeiner metric. Besides, it is found that the curvature divergent at the critical point of phase transition\cite{Ruppeiner:2011gm}.

For black hole systems, because there does not exist a complete theory of quantum gravity, although the most likely candidate theories---string theory and loop quantum gravity theory---have achieved good results to some extent, the exploration of the microscopic structure of black holes is bound to some speculative assumptions. Because of the development of black hole thermodynamics, we maybe speculate some micro-dynamics of the black hole from its thermodynamics. This is in some sense the reverse process of statistical physics, in which the macroscopic quantities are obtained from the microscopic state of the systems.
The idea of this process is reflected in the exploration of the underlying microscopic behavior of black holes by the Ruppeiner thermodynamic geometry.  Furthermore, it is still unclear about the constituents of black holes, hence the abstract concept of black hole molecule may be a good choice. Through analogy, we can imagine that there is an interaction among the molecules that make up the black hole. And we speculate that the empirical observation of the thermodynamic curvature corresponding to the interaction among the constituent molecules of the system also applies to black holes.

So the application of the Ruppeiner geometry into black hole thermodynamics makes it possible to glimpse the underlying microstructure of black holes from macroscopic thermodynamic quantities. With the help of Ruppeiner geometry, interesting features are observed in other black hole systems\cite{Sahay:2010wi,Niu:2011tb,Mansoori:2013pna,Chaturvedi:2014vpa,Sahay:2017hlq,Chaturvedi:2017vgq,Wei:2017icx,Cai:1998ep,Vacaru:2001ur,Aman:2003ug,Ruppeiner:2013yca,Dehyadegari:2016nkd,Li:2017xvi,KordZangeneh:2017lgs,Chen:2018icg,Miao:2018qyh,Guo:2019oad,Xu:2019nnp,Ghosh:2019pwy,Kumara:2019xgt}. Especially, in some situations that the entropy and thermodynamic volume of these black holes are not independent, such as Schwarzchild-AdS black hole, Reissner-Nordstr\"{o}m AdS (RN-AdS) black hole and etc., the heat capacity at constant volume of the black hole becomes zero. A vanishing heat capacity brings about a divergent thermodynamic curvature. Consequently, some micro information of the associated black hole is missing from the thermodynamic geometry approach. To cure this problem, either we can consider other suitable thermodynamic characteristic function as the fundamental starting point of Ruppenier geometry\cite{Xu:2019gqm,Ghosh:2020kba}, or introduce a normalized scalar curvature of thermodynamic geometry\cite{Wei:2019uqg}. In this paper, we adopt the latter. Novel thermodynamical features of black hole systems are observed with the help of normalized thermodynamic curvature, which shows the distinguishable behaviors from that of Van der Waals fluid\cite{Wei:2019yvs,Wei:2019ctz}.

In light of these rich phenomenologies, we are in anticipation of things to be more intriguing when the massive gravity model is considered. The first attempt to endow the graviton with mass is the work of Fierz and Pauli in the context of linear theory\cite{Fierz:1939ix}. Unfortunately, the predictions of Fierz-Pauli linear massive gravity theory in the massless limit do not coincide with those of general gravity, which is called van DamVeltman-Zakharov discontinuity\cite{vanDam:1970vg,Zakharov:1970cc}. Furthermore, the nonlinear level in a generic Fierz-Pauli theory will suffer the so-called Boulware-Deser ghost\cite{Boulware:1973my}. These seem to put an end to the massive gravity theory, until de Rham, Gabadadze and Tolley (dRGT)\cite{deRham:2010ik} put out a special class of nonlinear massive gravity that is ``Boulware-Deser'' ghost-free\cite{Hassan:2011hr}. It is found that in dRGT massive gravity theory, the effective cosmological constant can emerge from the massive parameters, rather than a fundamental quantity appearing in the action of the theory. The value of the effective cosmological constant is determined by the value of the massive parameters in the massive gravity model. In the case of the effective cosmological constant is positive, the massive gravity theory takes the advantage that provides a possible explanation for the accelerated expansion of the universe without any dark energy.

In addition, the dRGT massive gravity theory also admits the black hole solution in the case of a negative effective cosmological constant. In the literature \cite{Vegh:2013sk}, Vegh developed a nontrivial black hole solution in four dimensional massive gravity with a negative cosmological constant. He found that the mass of graviton behaves like a lattice excitation and exhibits a Drude peak in the dual field theory within the framework of gauge/gravity duality. The feature that massive gravity in AdS background introducing the momentum dissipation in the dual boundary field theory attract great interests in the application of gauge/gravity duality, such as the study on the temperature dependence of the shear viscosity to entropy density ratio \cite{Hartnoll:2016tri}, the holographic plasmon relaxation\cite{Baggioli:2019aqf}. Later, it was generalized to study the corresponding thermodynamical properties and phase structures\cite{Cai:2014znn,Xu:2015rfa}. More black hole solutions and their thermodynamical properties were addressed in arbitrary dimensional dRGT massive gravity theory with diverse corrections or matters\cite{Mirza:2014xxa,Hendi:2015bna,Do:2016abo,Li:2016fbf,Ning:2016usb}. One of the most prominent features that the thermodynamical behaviors of dRGT massive gravity have provided is that the so-called small/large black hole phase transition for the charged AdS black holes always exist, no matter the horizon topology is spherical $(k = 1)$, Ricci flat $(k = 0)$ or hyperbolic $(k =-1)$\cite{Ghosh:2015cva,Hendi:2017fxp}. This is the significant difference that the Van der Waals-like phase transition was usually only recovered in a variety of spherical horizon black hole backgrounds. Subsequently, the abundant thermodynamical behaviors of the black holes in dRGT massive gravity were investigated, such as the reentrant phase transitions,  tricritical point\cite{Zou:2016sab,Zhang:2017lhl,Liu:2018jld,Chaloshtary:2019qvv}. It is also suggested that the phase transition of the dRGT black holes could relate to the Quasinormal mode\cite{Prasia:2016fcc,Zou:2017juz} or the holographic entanglement entropy\cite{Zeng:2015tfj}.

In the literature\cite{Chabab:2019mlu}, the authors find that the singularities of scalar curvatures which are constructed from HPEM metric and the Gibbs free energy metrics coincide with the critical point of phase transition for the spherical black hole in dRGT massive gravity.

Motivated by these interesting results, in this paper, our aim is to investigate the Ruppeiner geometry with normalized scalar curvature proposed in \cite{Wei:2019uqg} and the phase transition of topological black holes in dRGT massive gravity. The outline of this paper is as follows. In section \ref{II}, we briefly review the Ruppeiner geometry, and derive the scalar curvature of black holes in the temperature and volume phase space $\{T,V\}$. In section \ref{III}, we study the thermodynamic properties of the topological black hole in dRGT massive gravity. The normalized scalar curvature and critical phenomena of these black holes are  investigated in section \ref{IV}. We end the paper with a summary and discussion in section \ref{V}. Throughout this paper, we adopt the units  $\hbar= c = k_B = G = 1$.

\section{\textsf{Ruppeiner Thermodynamic Geometry}}\label{II}

In this section, we are going to provide a brief introduction to the Ruppeiner thermodynamic geometry.
Thermodynamic geometry which originates from the fluctuation theory of equilibrium thermodynamics provides a possible way to explore the microscopic structure of black holes.

Consider an equilibrium isolated thermodynamic system with total entropy $S$, and
divide it into two subsystems of different sizes. One of these two sub-systems is small $S_B$, another one is a large subsystem $S_E$ and can be regarded as a thermo-bath. We additionally require that $S_B \ll S_E \sim S$. So the total entropy of the system describing by two independent thermodynamic
parameters $x^0$ and $x^1$ have the form of
\begin{equation}
    S(x^0,x^1)=S_B(x^0,x^1)+S_E(x^0,x^1).    \nonumber
\end{equation}
For a system in an equilibrium state, the entropy $S$ is at its local maximum value. Expanding the total entropy at the vicinity of the local maximum ($x^\mu=x^\mu_0$), we have
\begin{align}
    S=&S_0+\frac{\partial S_B}{\partial x^\mu}\bigg|_{x^\mu_0}\Delta x^\mu_B
          +\frac{\partial S_E}{\partial x^\mu}\bigg|_{x^\mu_0}\Delta x^\mu_E    \nonumber\\
      &+\frac{1}{2}\frac{\partial^2 S_B}{\partial x^\mu \partial x^\nu}\bigg|_{x^\mu_0}\Delta x^\mu_B \Delta x^\nu_B
       +\frac{1}{2}\frac{\partial^2 S_E}{\partial x^\mu \partial x^\nu}\bigg|_{x^\mu_0}\Delta x^\mu_E \Delta x^\nu_E
       +\cdots,    \nonumber
\end{align}
here $S_0$ is the local maximum at $x^0_\mu$. The entropy of the equilibrium isolated
system is conserved under the virtual change which indicates that the first derivative
of it is zero, so we promptly arrive at
\begin{align}
    \Delta S=S-S_0&=\frac{1}{2}\frac{\partial^2 S_B}{\partial x^\mu \partial x^\nu}\bigg|_{x^\mu_0}\Delta x^\mu_B \Delta x^\nu_B
       +\frac{1}{2}\frac{\partial^2 S_E}{\partial x^\mu \partial x^\nu}\bigg|_{x^\mu_0}\Delta x^\mu_E \Delta x^\nu_E
       +\cdots    \nonumber  \\
       &\approx \frac{1}{2}\frac{\partial^2 S_B}{\partial x^\mu \partial x^\nu}\bigg|_{x^\mu_0}\Delta x^\mu_B \Delta x^\nu_B.
\end{align}
Since the entropy $S_E$ of the bath as an extensive thermodynamical quantity gets the same order as that of the whole system, so its second derivatives with respect to the intensive thermodynamical quantities $x^\mu$ are much smaller than those of $S_B$ and has been ignored in the second line.

In Ruppeiner geometry, the entropy is treated as the thermodynamical potential and its fluctuation $\Delta S$
is associated with the line element $\Delta l^2$ with thermodynamical coordinates $\{x^\mu\}$\cite{Ruppeiner:1995zz}. The Ruppeiner line element reads as
\begin{align}
    \Delta l^2=&\frac{1}{k_B}g^R_{\mu \nu} \Delta x^\mu \Delta x^\nu,
\end{align}
where $k_B$ is the Boltzmann constant and the Ruppeiner metric $g^R_{\mu \nu}$ is
\begin{align}
	g^R_{\mu \nu}&=-\frac{\partial^2 S_B}{\partial x^\mu \partial x^\nu}.
\end{align}
Here $\Delta l^2$ measures the distance between two neighboring fluctuation states. Thus thermodynamic metric $g^R_{\mu \nu}$ has a large potential to shine some lights on the information about microstructure of systems.

Now we set the thermodynamical coordinates ${x^\mu}$ as temperature $T$ and volume
$V$, and take the Helmholtz free energy as the thermodynamical potential. The line element $\Delta l^2$ is given by \cite{Wei:2019uqg}
\begin{equation}
    \Delta l^2=\frac{C_V}{T^2} \Delta T^2-\frac{(\partial_V P)_T}{T} \Delta V^2,    \label{IM}
\end{equation}
where $C_V=T(\partial_T S)_V$ is the heat capacity at constant volume.
Adopting the convention used in \cite{Ruppeiner:1995zz}, the scalar curvature of the above line element can be worked out directly, and it yields
\begin{align}
	R&=\frac{1}{2C_V^2(\partial_V P)^2}\bigg \{ T(\partial_V P) \Big[ (\partial_T C_V)(\partial_V P-T\partial_{T,V} P) + (\partial_V C_V)^2	\Big]  \nonumber \\
	& +C_V \Big[ (\partial_V P)^2+ T((\partial_V C_V)(\partial^2_V P)-T(\partial_{T,V} P)^2)+2T(\partial_V P)(T(\partial_{T,T,V} P)-(\partial^2_V C_V))
	\Big] \bigg \}.    \label{scalarCurva}
\end{align}
It is believed that the sign of the scalar curvature encodes the effective interaction between two microscopic molecules of a given system, i.e., $R >0$ and $R <0$ are associated with repulsive and attractive interaction, respectively. While $R=0$ implies that the repulsive and attractive interaction are in balance\cite{Ruppeiner:1995zz}. Moreover, it is speculated that the divergence of the scalar curvature occurs at the critical point of phase transition.
Although, because of the absence of a hitherto underlying theory of quantum gravity, there has been a controversial discussion on whether the characterizations of thermodynamic scalar curvature can be directly generalized to the black hole thermodynamics or not\cite{Dolan:2015xta}.  But the well-established black hole thermodynamics makes the Ruppeiner thermodynamic geometry be plausible to phenomenologically or qualitatively provide the information about interactions of black holes.

\section{\textsf{Thermodynamics of dRGT massive gravity}}\label{III}
In this section, we would like to briefly review the topological dRGT black hole, and discuss the thermodynamic properties of it in extended phase space. The action describing a $(n+2)$-dimensional charged topology AdS
black hole in dRGT massive gravity\cite{Vegh:2013sk} is
\begin{equation}
    I=\frac{1}{16\pi G}\int \textrm{d}^{n+2} x \sqrt{-g} \left(R-2 \Lambda-\frac{1}{4} F_{\mu \nu} F^{\mu \nu}
        +m^2 \sum \limits^{4}_{i=1} c_i {\cal U}_i(g,f)\right),    \label{Q1}
\end{equation}
in which $\Lambda$ is the cosmology constant, $F_{\mu \nu}$ is the electromagnetic field tensor
defined as $F_{\mu \nu} = \partial_{\mu} A_{\nu}-\partial_{\nu}A_{\mu}$ with vector potential $A_{\mu}$.
Moreover, $f$ is the reference metric coupled to the
metric $g_{\mu \nu}$. The mass of the graviton is related to the parameter $m$,
$c_i$'s are dimensionless constants, and ${\cal U}_i$'s are symmetric polynomials constructed from the $(n+2) \times (n+2)$ matrix ${\cal K}^\mu {}_{\nu}=\sqrt{g^{\mu \alpha}f_{\nu \alpha}}$, they take the forms of
\begin{align*}
{\cal U}_{1}&=[\cal K],\nonumber\\
{\cal U}_{2}&=[{\cal K}]^2-[{\cal K}^2],\nonumber\\
{\cal U}_{3}&=[{\cal K}]^3-3[{\cal K}][{\cal K}^2]+2[{\cal K}^3],\nonumber\\
{\cal U}_{4}&=[{\cal K}]^4-6[{\cal K}^2][{\cal K}]^2
+8[{\cal K}^3][{\cal K}]+3[{\cal K}^2]^2-6[{\cal K}^4].
\end{align*}
The square root in ${\cal K}$ represents $(\sqrt{{\cal K}})^\mu{}_{\nu} (\sqrt{{\cal K}})^\nu{}_{\lambda}={\cal K}^\mu{}_\lambda$.

The topological black hole solution allows the following form of metric
\begin{equation}
    \mathrm{d} s^2=-f(r)\mathrm{d} t^2+g(r)\mathrm{d} r^2+r^2 \gamma_{ij} \mathrm{d} x^i \mathrm{d} x^j,    \label{Q2}
\end{equation}
where $\gamma_{ij}$ is the metric of a $n$-dimensional hypersurface with constant
scalar curvature $n(n-1)k$. In general, the value of $k$ can be $1, 0$ or $-1$
corresponding to spherical, planar and hyperbolic topology, respectively.
With a special choice of reference metric in \cite{Vegh:2013sk}
\begin{equation}
    f_{\mu \nu}={\rm diag}  (0,0, c^2_0 \gamma_{ij}),    \label{Q3}
\end{equation}
${\cal U}_i$'s can be obtained as
\begin{align}
{\cal U}_{1}&=nc_0/r,\nonumber\\
{\cal U}_{2}&=n(n-1)c^{2}_{0}/r^2,\nonumber\\
{\cal U}_{3}&=n(n-1)(n-2)c^{3}_{0}/r^3,\nonumber\\
{\cal U}_{4}&=n(n-1)(n-2)(n-3)c^{4}_{0}/r^4.
\end{align}
In the four-dimensional spacetime ($n=2$), it is easily to verify that ${\cal U}_3 = {\cal U}_4 = 0$, which prevents $c_3$ and $c_4$ from appearing in the spacetime metric. It indicate that these two parameters have
no effect on the phase structure of the four-dimensional topological black hole in massive gravity.
Then the metric function has the form of \cite{Cai:2014znn}
\begin{equation}
    f(r)=\frac{1}{g(r)}=k-\frac{\Lambda r^2}{3}-\frac{m_0}{r}+\frac{q^2}{4r^2}
         +\frac{c_0 c_1 m^2}{2} r+c^2_0 c_2 m^2.    \label{Q4}
\end{equation}
Here the parameters $m_0$ and $q$ relates to the mass and charge amount of the black hole
\begin{equation}
    M=\frac{\omega_2}{8 \pi} m_0, \quad Q=\frac{\omega_2}{16 \pi} q.
\end{equation}
in which $\omega_2$\footnote{For a spherical topology with $k=1$, the volume of the two-dimensional hypersurface is $\omega_2=4\pi$; when $k=0$ or $-1$, with the different compaction methods, $\omega_2$ take different positive numbers \cite{Ong:2015fha}.} is the volume of the two-dimensional hypersurface. Without loss of generality, we will set $c_0 = 1$ and $m=1$ in our following discussions, but leave $c_1$ and $c_2$ as free parameters.
In the extended phase space, the negative cosmological constant is treated as the pressure of the black hole $P=-\Lambda/8 \pi$, which leads to an interpretation of the black hole mass as enthalpy. In terms of the definition of pressure $P$ and the radius of the horizon $r_0$, the thermodynamics quantities of the black hole, i.e., mass $M$, temperature $T$, and electromagnetic potential $\Phi$ can be obtained as in \cite{Xu:2015rfa}
\begin{align} \label{thermo}
	M&=\frac{\omega_2 r_0}{8\pi}\left(k+ c_2+\frac{q^2}{4r_0^2}+\frac{8\pi Pr_0^2}{3}+\frac{c_1 r_0}{2}
	\right) \nonumber \\
	T&=2r_0P+\frac{k+c_2}{4\pi r_0}-\frac{q^2}{16\pi r_0^3}+\frac{c_1}{4\pi}, \nonumber \\
	S&=\frac{\omega_2 }{4} r_0^2, \qquad \Phi=\frac{q}{r_0}.
\end{align}
We can also easily get the thermodynamic volume conjugated to the pressure\cite{Kubiznak:2012wp}
\begin{equation}
    V=\left(\frac{\partial M}{\partial P}\right)_{S,Q}=\frac{4\pi}{3}r_0^3= \frac{\pi}{6}v^3.    \label{Q6}
\end{equation}
where a specific volume \cite{Kubiznak:2012wp,Toledo:2019amt} defined as $v=2r_0$ we introduced is not the normal definition of specific volume in normal thermodynamic (volume per unit mass). In what following, we would see that the definition of the specific volume $v=2r_0$ for the black hole is a naturally choice comparing to the equation of state of the VdW fluid.

From Eqs. (\ref{thermo}) and (\ref{Q6}), the equation of state of dRGT black holes turns to\cite{Xu:2015rfa}
\begin{equation}
    P=\frac{T}{v}-\frac{c_1}{4\pi v}-\frac{k+c_2}{2\pi v^2}+\frac{q^2}{2\pi v^4},     \label{Q12}
\end{equation}
which is analog to that of Van der Waals fluid. The thermodynamics theory
tells us that the critical point of second order phase transition located at the inflection point of pressure, the condition is
\begin{equation}
    \left(\frac{\partial P}{\partial v}\right)_T=0, \quad \left(\frac{\partial^2 P}{\partial v^2}\right)_T=0.    \label{critical}
\end{equation}
It leads to the following critical quantities\cite{Xu:2015rfa}
\begin{align}
     T_c=\frac{c_1}{4\pi}+\frac{2 (k+c_2)^{3/2}}{3\sqrt{6}\pi |q|},
     \quad P_c=\frac{(k+c_2)^{2}}{24\pi |q|^2},
     \quad v_c=\frac{\sqrt{6} |q|}{\sqrt{k+c_2}}.    \label{cp}
\end{align}
The critical point of the black hole phase transition solved from Eq.(\ref{critical}) contains the absolute value of charge $|q|$, which implies that the result we obtained is symmetric to the sign of the charge. For convenience, we use the symbol $q$ to replace $|q|$ in the following discussion. It is worth noticing from Eq.(\ref{cp}) that usually the critical phenomenon of black hole only recovered in a variety of spherical topology ($k=1$) in previous works. However, from the above critical quantities, we find that all values of $k$ $(0,\pm 1)$ have the possibility to admit Van der Waals- like behaviors provided $k+c_2>0$. This is the significant feature of dRGT massive gravity, that a non-zero parameter $m$ related to the gravity mass greatly enriches the thermodynamics behaviors of topological black holes.

One can easily verify that in the absence of massive parameters $c_1$ and $c_2$, i.e., $c_1=0=c_2$,
the critical coefficient defined as the ratio of critical temperature $T_c$ to the multiplier of critical pressure and volume $P_c v_c$ equals to
\begin{equation}
    \frac{T_c}{P_c v_c}=\frac{8}{3},
\end{equation}
which coincides with the case of the normal Van der Waals fluid.
However, in the presence of massive parameters $c_1$ and $c_2$, the critical coefficient becomes
\begin{equation}
    \frac{T_c}{P_c v_c}=\frac{8}{3} + \frac{\sqrt{6}c_1 q}{(k+c_2)^{3/2}},
\end{equation}
which receives a correction combined by the parameters $c_1, q, k, c_2$. In the next subsection, we focus on the relation between the Rupperiner geometry with normalized scalar curvature and the phase transition of topological black holes in dRGT massive gravity.

\subsection{\textsf{Phase transition of Charged dRGT black hole}}

In this subsection, the thermodynamical characteristics of charged dRGT black hole with $k+c_2>0$ which undergoes phase transition are described, and we further explore how the metic parameters affect the thermodynamic properties of black holes. We first introduce the reduced parameters defined as
\begin{equation}
    \widetilde{P}=\frac{P}{P_c},
    \quad \widetilde{T}=\frac{T}{T_c},
    \quad \widetilde{v}=\frac{v}{v_c},
    \quad \widetilde{S}=\frac{S}{S_c}.    \nonumber
\end{equation}
Then the equation of state Eq.(\ref{Q12}) of the black hole arrives at
\begin{equation}
    \widetilde{P}=\frac{8}{3 \tilde{v}} \left\{ \widetilde{T} \left[ 1+\frac{c_1 q}{16}
                   \left( \frac{6}{k+c_2} \right)^\frac{3}{2} \right]
                   -\frac{c_1 q}{16} \left( \frac{6}{k+c_2}\right)^\frac{3}{2}\right\}
                   -\frac{2}{\tilde{v}^2}+\frac{1}{3\tilde{v}^4}.   \label{pv}
\end{equation}
A small/large black hole phase transition occurs when $\widetilde{T}<1$, this first order phase transition can be identified from Maxwell's equal area law\cite{Spallucci:2013osa}. It will more suitable for us to verify the equal area law in $(\widetilde{T},\widetilde{S})$ phase space. After considering the relation of $S$ and $v$, the equation of state described by the reduced parameters turns to
\begin{equation}
    \widetilde{T}\left[1+\frac{c_1 q}{16} \left(\frac{6}{k+c_2}\right)^\frac{3}{2}\right]
                -\frac{c_1 q}{16}\left(\frac{6}{k+c_2}\right)^\frac{3}{2}
                =\frac{3}{4\sqrt{\widetilde{S}}}\left(\frac{1}{2}\widetilde{P}\widetilde{S}
                    +1-\frac{1}{6\widetilde{S}}\right).    \label{Q8}
\end{equation}
Here we notice that there is a combination of parameters $c_1, q, k, c_2$, which takes a crucial role in the equation of state. So it is necessary for us to introduce a combined parameter $x=\frac{c_1 q}{16}\left(\frac{6}{k+c_2}\right)^\frac{3}{2}$ and investigate its effects on the phase structure of the massive black hole.

With Eq.(\ref{pv}), the isothermal curves $\widetilde{T}=0.8$
are plotted in Fig.\ref{Fig1}(a) in reduced $(\widetilde{P},\tilde{v})$ phase space from bottom to top with $x=0.3$, $0$ and $-0.3$. As shown in the figure, there are two extreme points on the isotherms, and the curves are divided into three parts. It can be seen that as $x$ decreases, the two extreme points keep approaching.
\begin{figure}[!h]
\centering
\subfigure[]{
\includegraphics[width=7cm]{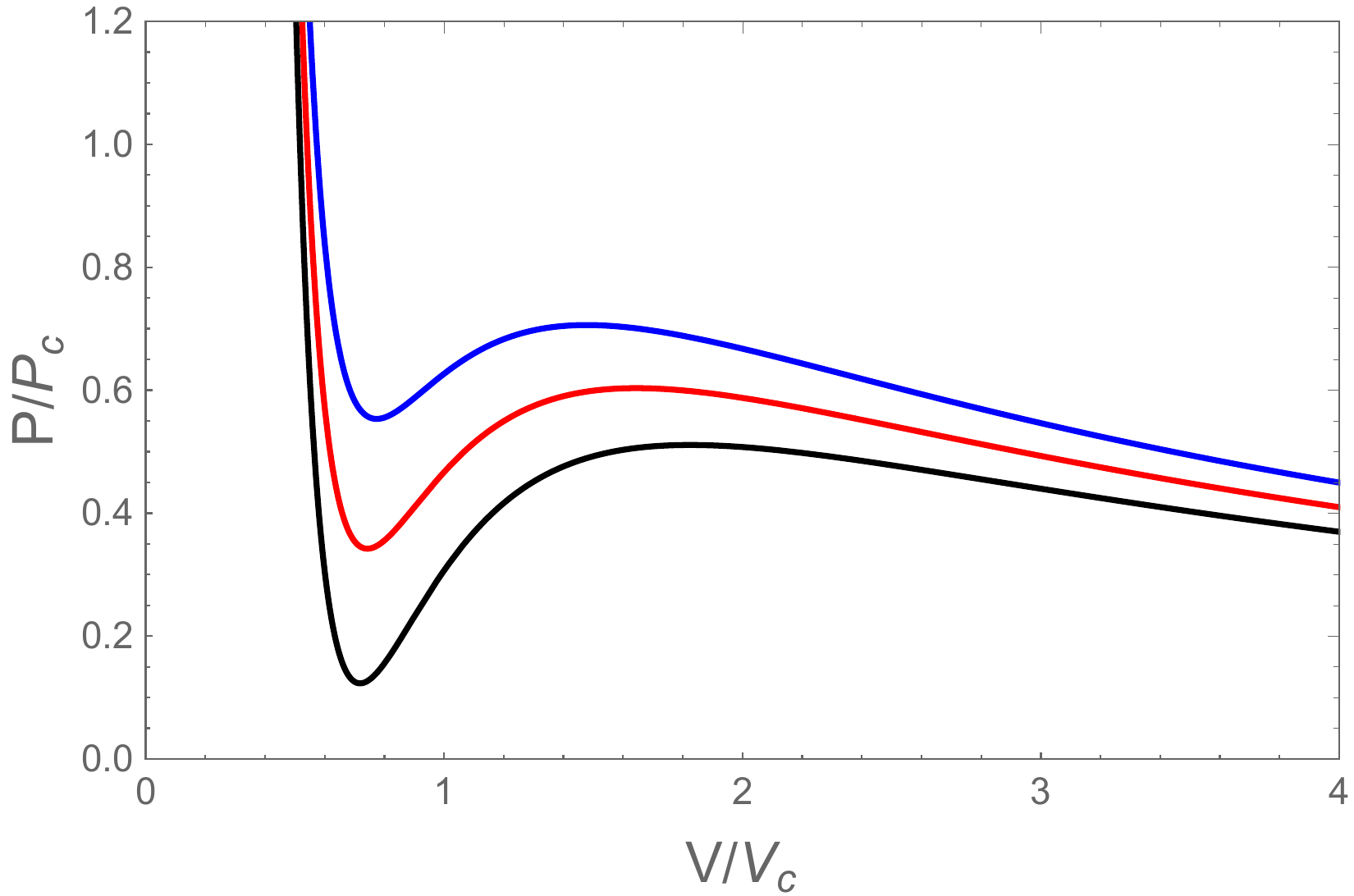}
 }
\quad
\subfigure[]{
\includegraphics[width=7cm]{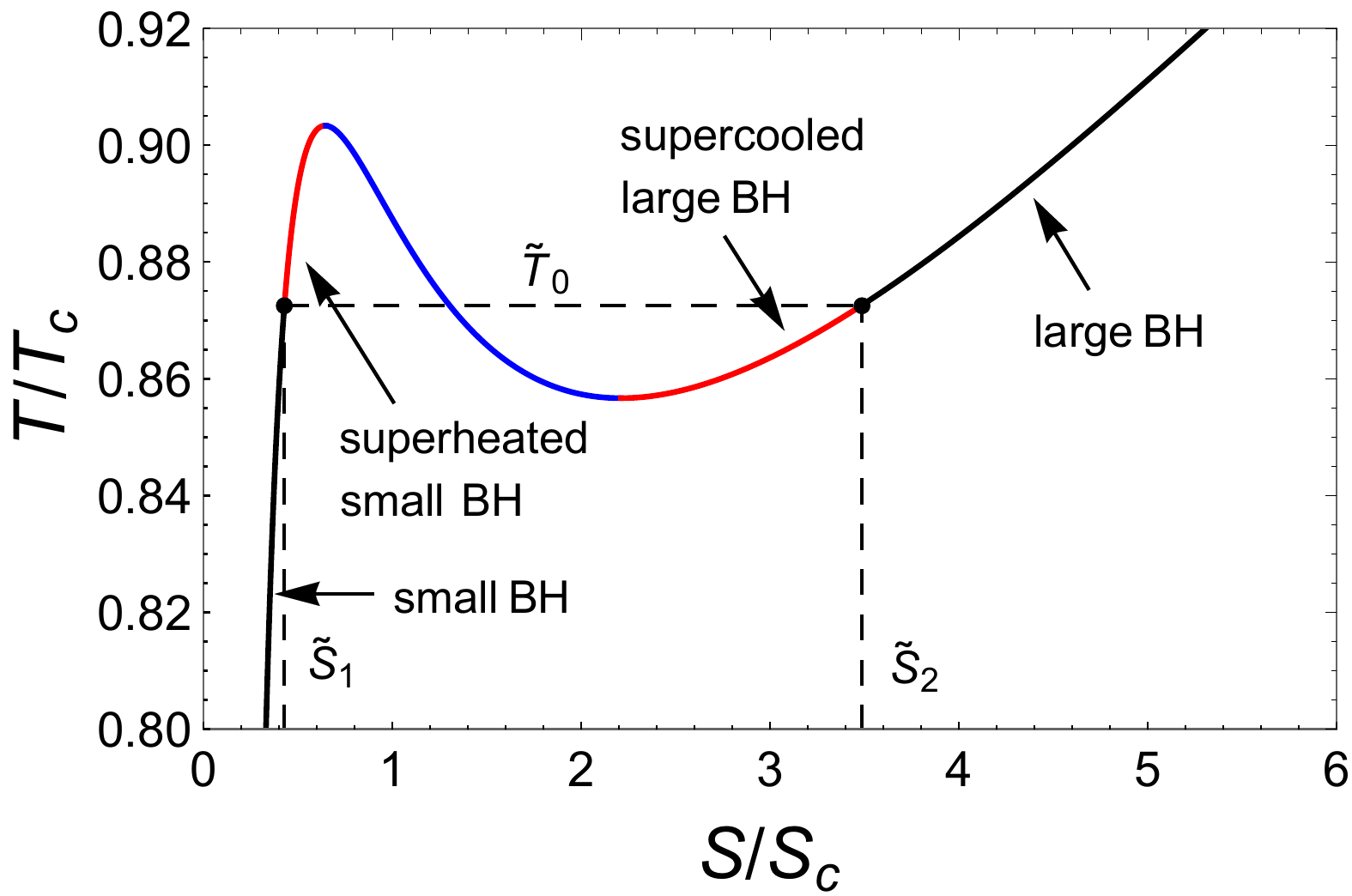}
}
\caption{(a) Isotherms curves with $x=0.3$, $0$, $-0.3$ and $\widetilde{T}=0.8$ from bottom to top in reduced $(\widetilde{P},\tilde{v})$ space. (b) The isobar curve in reduced $(\widetilde{T},\widetilde{S})$ plane with $\widetilde{P}=0.7$ and $x=0$. Black solid curves describe the small black holes and large black holes. The red curves represent the superheated small black hole and the supercooled large black hole. And the blue curve indicates the black hole is in an unstable phase. The black dashed line implied by equal area law is the phase transition temperature.}\label{Fig1}
\end{figure}
With Eq.(\ref{Q8}), we construct an isobar curve in reduced $(\widetilde{T},\widetilde{S})$ phase space with $\widetilde{P}=0.7$ and $x=0$  for example. The isobar curve is shown in Fig.\ref{Fig1}(b) is correspondingly divided into three pieces by two extreme points. Black solid curves represent small black holes phase and large black holes phase. The red curves represent the superheated small black hole phase and supercooled large black hole phase. And the blue curve whose slope is negative indicates the black hole is in an unstable phase. The extreme points separating the metastable phase from the unstable phase are called spinodal points. The black horizontal dashed line determined by the Maxwell equal area law is the phase transition temperature $\widetilde{T}_0$, which indicates the area under the $\widetilde{T}(\widetilde{S})$ curve from $\widetilde{S}_1$ to $\widetilde{S}_2$ is the same as the area enclosed by the black dashed line and coordinate axis. Thus we get the following conditions
\begin{align}
    \widetilde{T}_0 (\widetilde{S}_2&-\widetilde{S}_1)
        =\int_{\widetilde{S}_1}^{\widetilde{S}_2} \widetilde{T} (\widetilde{S}){\rm d}\widetilde{S}, \, \qquad
    \widetilde{T}_0=\widetilde{T} (\widetilde{S}_1)=\widetilde{T} (\widetilde{S}_2),
\end{align}
from which we get the solutions $\widetilde{S}_1$, $\widetilde{S}_2$, and $\widetilde{T}_0$, read as
\begin{align}
    \sqrt{\widetilde{S}_{1,2}}=&\frac{ \sqrt{ 3-\sqrt{\widetilde{P}} } \mp  \sqrt{3-3\sqrt{\widetilde{P}}}}
                                     { \sqrt{ 2 \widetilde{P}}},    \label{entropy}\\
    \widetilde{T}_{0}\left[1+\frac{c_1 q}{16}\left(\frac{6}{k+c_2}\right)^\frac{3}{2}\right]
               &-\frac{c_1 q}{16}\left(\frac{6}{k+c_2}\right)^\frac{3}{2}
                   =\sqrt{\frac{\widetilde{P}\left(3-\sqrt{\widetilde{P}}\right)}{2}}.    \label{Q7}
\end{align}
The two particular points $(\widetilde{T}_{0},\widetilde{S}_{1})$ and $(\widetilde{T}_{0},\widetilde{S}_{2})$ in Fig.\ref{Fig1}(b) describe respectively the saturated small black hole and the saturated large black hole. The Eq.(\ref{Q7}) is served as the small/large black hole coexistence curve, which can be inversed to
\begin{equation}
    \sqrt{\widetilde{P}}=1-2\sin \left\{\frac{1}{3} \arcsin \left\{1-\left\{\widetilde{T}_0 \left[1+\frac{c_1 q}{16}
        \left( \frac{6}{k+c_2} \right)^\frac{3}{2} \right]
        -\frac{ c_1 q}{16} \left(\frac{6}{k+c_2} \right)^\frac{3}{2} \right\}^2 \right\} \right\}.    \label{co1}
\end{equation}
Substituting the equation of state Eq.(\ref{pv}) and the relation $\widetilde{V}=\tilde{v}^3$ into Eq.(\ref{Q7}), the form of the coexistence curve can be expressed as
\begin{equation}
    \widetilde{T}_{0}=
                  \left[\frac{5-3\sqrt{6\widetilde{V}^{\frac{2}{3}}+3}+6\widetilde{V}^{\frac{2}{3}}}{2\widetilde{V}}
                        +\frac{c_1 q}{16}\left(\frac{6}{k+c_2}\right)^\frac{3}{2}\right]
                         \Bigg /\left[1+\frac{c_1 q}{16} \left(\frac{6}{k+c_2}\right)^\frac{3}{2}\right].    \label{Q10}
\end{equation}
According to the relation between $V$ and $S$, we can get the thermodynamic volumes of the black hole corresponding to $\widetilde{S}_1$ and $\widetilde{S}_2$ on the isobar curve
\begin{align}
    \widetilde{V}_1&=\left(\frac{ \sqrt{ 3-\sqrt{\widetilde{P}} } -\sqrt{3-3\sqrt{\widetilde{P}}}}
                                     { \sqrt{ 2 \widetilde{P}}}\right)^3,    \label{V1}\\
    \widetilde{V}_2&=\left(\frac{ \sqrt{ 3-\sqrt{\widetilde{P}} } +\sqrt{3-3\sqrt{\widetilde{P}}}}
                                     { \sqrt{ 2 \widetilde{P}}}\right)^3.    \label{V2}
\end{align}
The change $\Delta\widetilde{V}=\widetilde{V}_2-\widetilde{V}_1$ is an extremely interesting parameter in the phase transition which can be treated as an order parameter. In the light of Eq.(\ref{V1}) and Eq.(\ref{V2}), together with Eq.(\ref{co1}), we plot the figures about the change $\Delta\widetilde{V}$ as functions of $\widetilde{P}$ and $\widetilde{T}$ respectively in Fig.\ref{Fig3}.
\begin{figure}[htbp]
\centering
\subfigure[]{
\includegraphics[width=7cm]{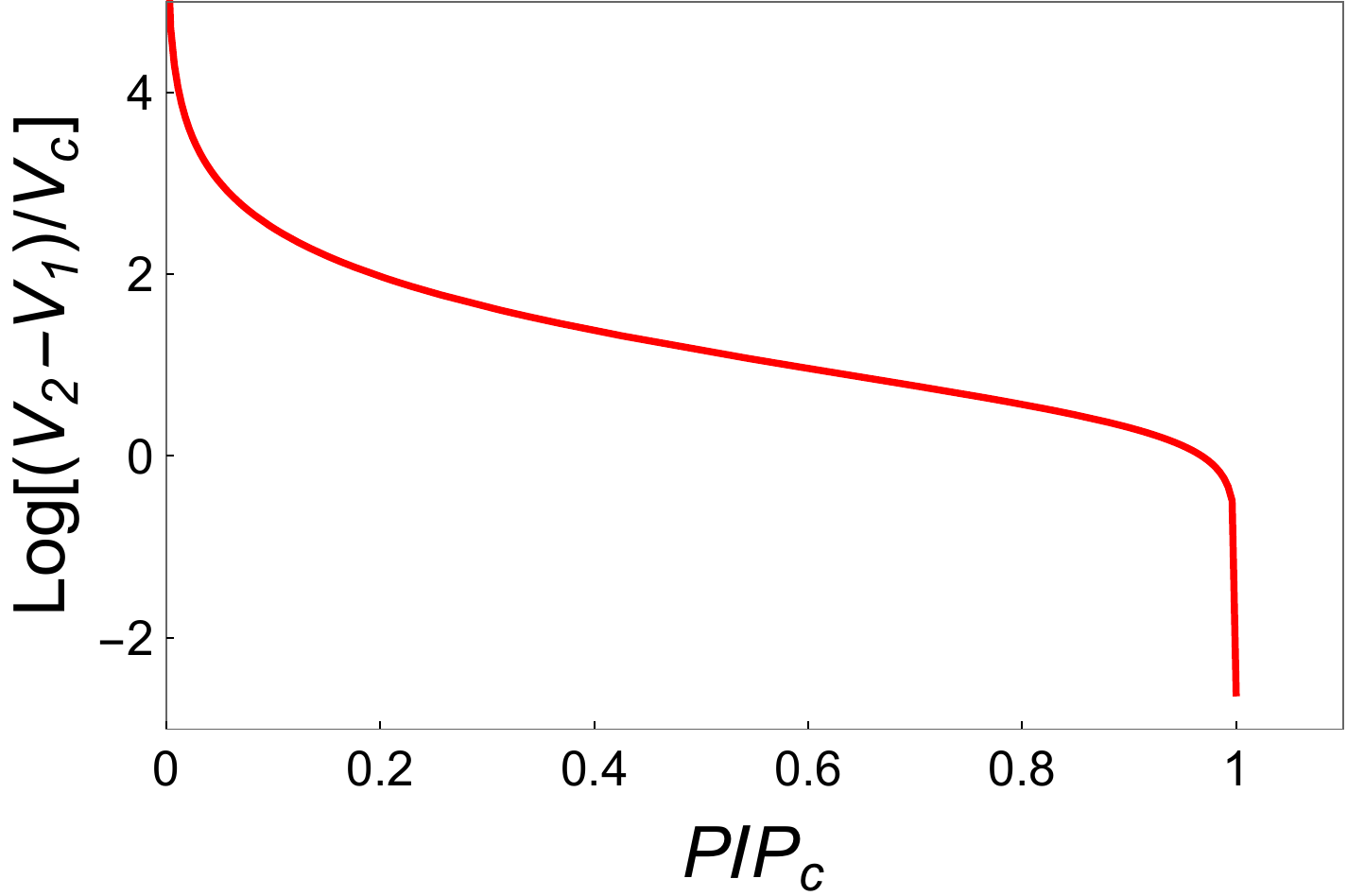}
}
\quad
\subfigure[]{
\includegraphics[width=7cm]{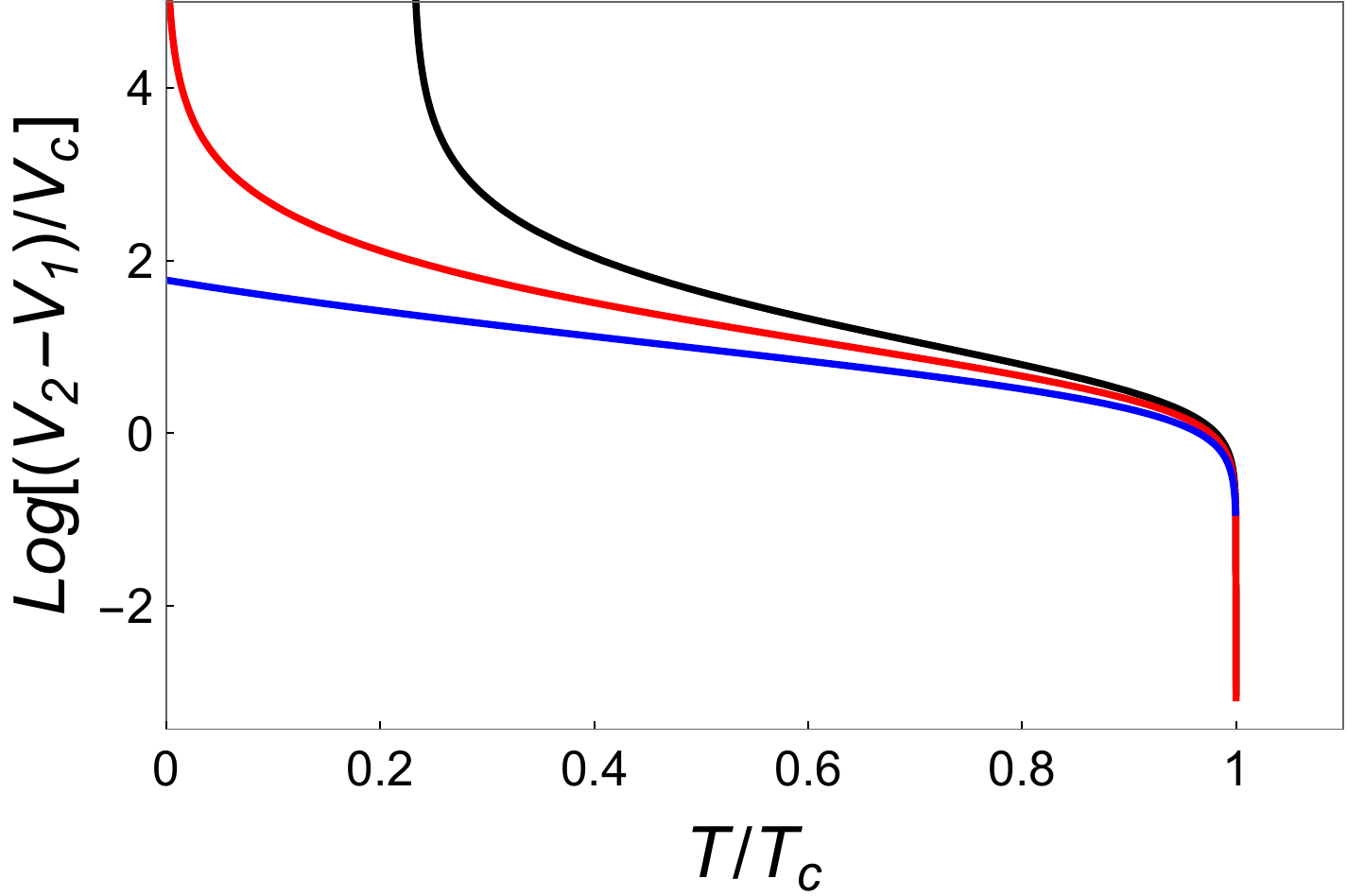}
}
\caption{(a) $\Delta\widetilde{V}$ as a function of $\widetilde{P}$ is independent of the parameter $x$. (b) $\Delta\widetilde{V}$ as a function of $\widetilde{T}$ with $x=0.3$, $0$ and $-0.3$ from top to bottom.}\label{Fig3}
\end{figure}
These two figures indicate that $\Delta\widetilde{V}$ decreases monotonically as $\widetilde{P}$ and $\widetilde{T}$ increase, and finally becomes zero at the critical point. The situation of $\Delta\widetilde{V}$ as a function of $\widetilde{T}$ will be more subtle. Fig.\ref{Fig3}(b) is the curve of $\Delta\widetilde{V}$ as a function of $\widetilde{T}$ with $x=0.3$, $0$ and $-0.3$ from top to bottom. As $x$ decreases the corresponding $\Delta\widetilde{V}$ decreases. The smaller $\Delta\widetilde{V}$ indicates the difference between the small black hole phase and the large black hole phase is smaller, the coexistence region in phase space is compressed. What's more, for positive parameter $x=0.3$, we can see from the black curve in Fig.\ref{Fig3}(b) that the $\Delta\widetilde{V}$ reaches infinity when the temperature $\widetilde{T}\le3/13$, which indicates that the difference between the saturated small black hole and the saturated large black hole is huge. It is an intriguing feature of dRGT black hole that we will further discuss in what follows. After performing series expansion on $\Delta\widetilde{V}$ at the critical point, we have
\begin{align*}
    \Delta\widetilde{V}&=3\sqrt{3}(1-\widetilde{P})^\frac{1}{2}
                         +\frac{51\sqrt{3}}{8}(1-\widetilde{P})^\frac{3}{2}
                         +\mathcal{O}(1-\widetilde{P})^\frac{5}{2},    \\
    \Delta\widetilde{V}&=6\sqrt{2+\frac{c_1 q}{8} \left(\frac{6}{k+c_2}\right)^\frac{3}{2}}
                         (1-\widetilde{T})^\frac{1}{2}
                        +\frac{61\sqrt{2}}{2}\left[1+\frac{c_1 q}{16} \left(\frac{6}{k+c_2}\right)^\frac{3}{2}\right]^\frac{3}{2} (1-\widetilde{T})^\frac{3}{2}
                         +\mathcal{O}(1-\widetilde{T})^\frac{5}{2}.    \nonumber
\end{align*}
Obviously, it can be seen that, like the RN-AdS black hole, the critical exponent of the dRGT massive black hole also takes the universal value of $1/2$.

Now let's turn our attention to the unstable phase describe by the solid blue curve in Fig.\ref{Fig1}(b). The spinodal curve separating the unstable phase and metastable phase is obtained from the condition $(\partial_V P)_T=0$, one gets
\begin{equation}
    \widetilde{T}_{\mathrm{sp}}=
        \left[\frac{3\widetilde{V}^{-\frac{1}{3}}-\widetilde{V}^{-1}}{2}
              +\frac{c_1 q}{16}\left(\frac{6}{k+c_2}\right)^\frac{3}{2}\right]
              \Bigg / \left[1+\frac{c_1 q}{16}\left(\frac{6}{k+c_2}\right)^\frac{3}{2}\right].    \label{Tsp}
\end{equation}
Inversing it, we can obtain the expression of the spinodal curves
\begin{equation}
    \widetilde{V}_1^\frac{1}{3}=\frac{1-2\cos\left\{\frac{1}{3}\arccos\left\{1-2\left\{\widetilde{T}
        \left[1+\frac{c_1q}{16}\left(\frac{6}{k+c_2}\right)^\frac{3}{2}\right]
        -\frac{c_1q}{16}\left(\frac{6}{k+c_2}\right)^\frac{3}{2}\right\}^2\right\}
        +\frac{\pi}{3}\right\}}
        {2\widetilde{T}\left[1+\frac{c_1q}{16}\left(\frac{6}{k+c_2}\right)^\frac{3}{2}\right]
        -\frac{c_1q}{8}\left(\frac{6}{k+c_2}\right)^\frac{3}{2}},    \label{V11}
\end{equation}
\begin{equation}
    \widetilde{V}_2^\frac{1}{3}=\frac{1+2\cos\left\{\frac{1}{3}\arccos\left\{1-2\left\{\widetilde{T}
        \left[1+\frac{c_1q}{16}\left(\frac{6}{k+c_2}\right)^\frac{3}{2}\right]
        -\frac{c_1q}{16}\left(\frac{6}{k+c_2}\right)^\frac{3}{2}\right\}^2\right\}\right\}}
        {2\widetilde{T}\left[1+\frac{c_1q}{16}\left(\frac{6}{k+c_2}\right)^\frac{3}{2}\right]
        -\frac{c_1q}{8}\left(\frac{6}{k+c_2}\right)^\frac{3}{2}},    \label{V22}
\end{equation}
where $\widetilde{V}_1<1$ is for the small black hole spinodal curve, while $\widetilde{V}_2>1$ is for the large black hole spinodal curve.

We plot the coexistence curve and spinodal curve in Fig.\ref{Fig4} in the reduced $(\widetilde{P},\widetilde{T})$ phase space with the red solid and blue dashed curves, respectively.
\begin{figure}[!h]
\begin{center}
\subfigure[$x=0.3$]{
\includegraphics[width=5cm]{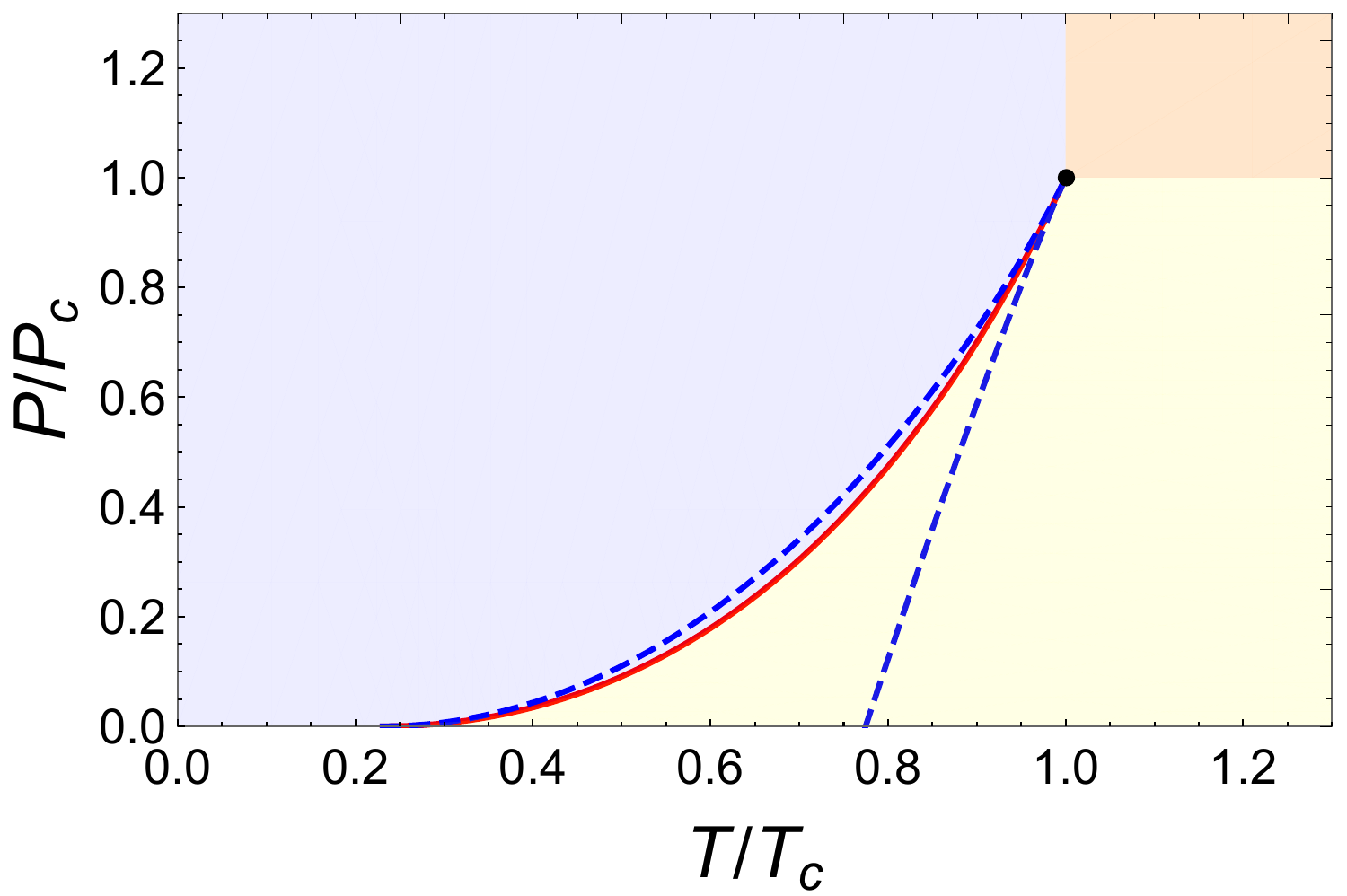}
}
\quad
\subfigure[$x=0$]{
\includegraphics[width=5cm]{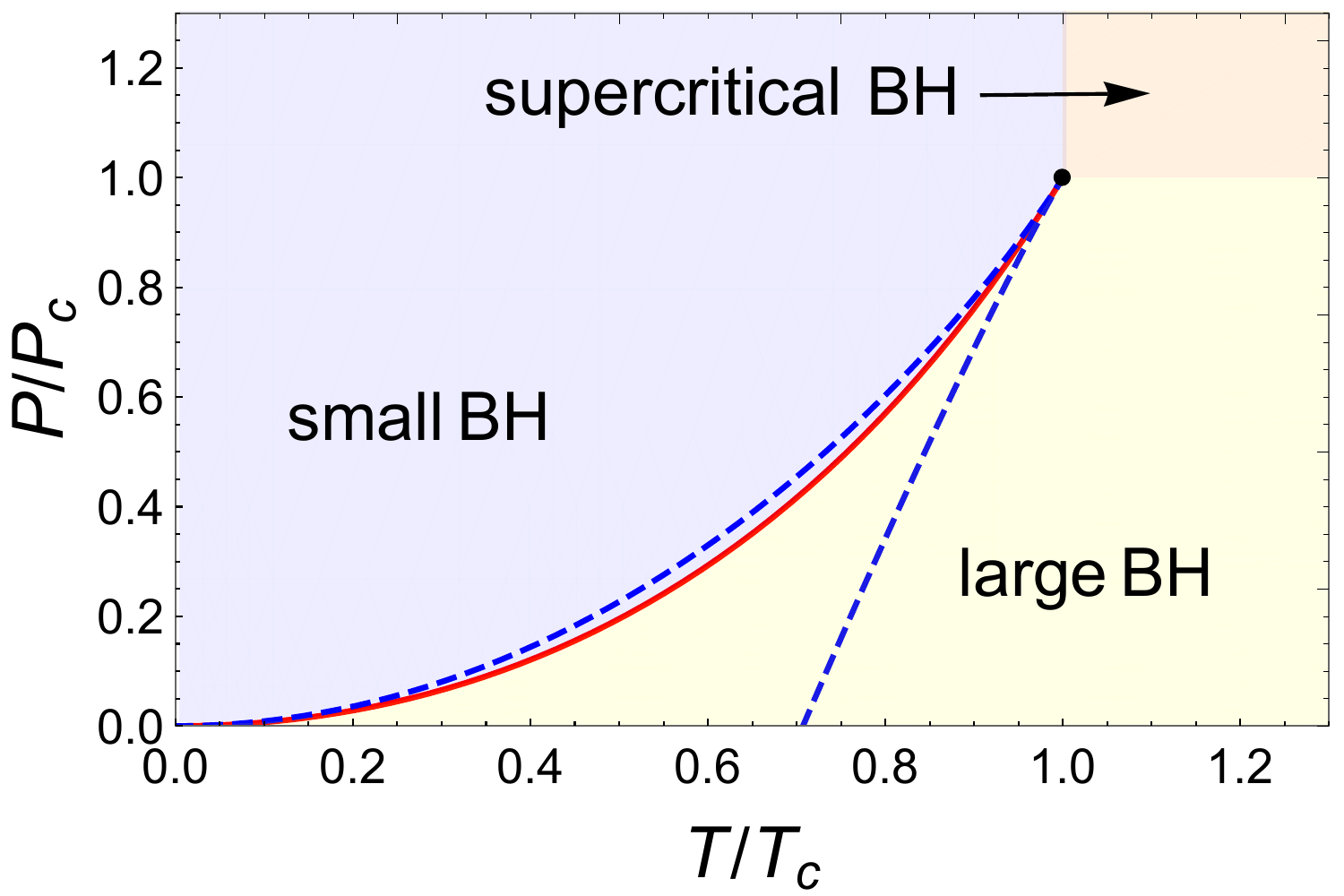}
}
\quad
\subfigure[$x=-0.3$]{
\includegraphics[width=5cm]{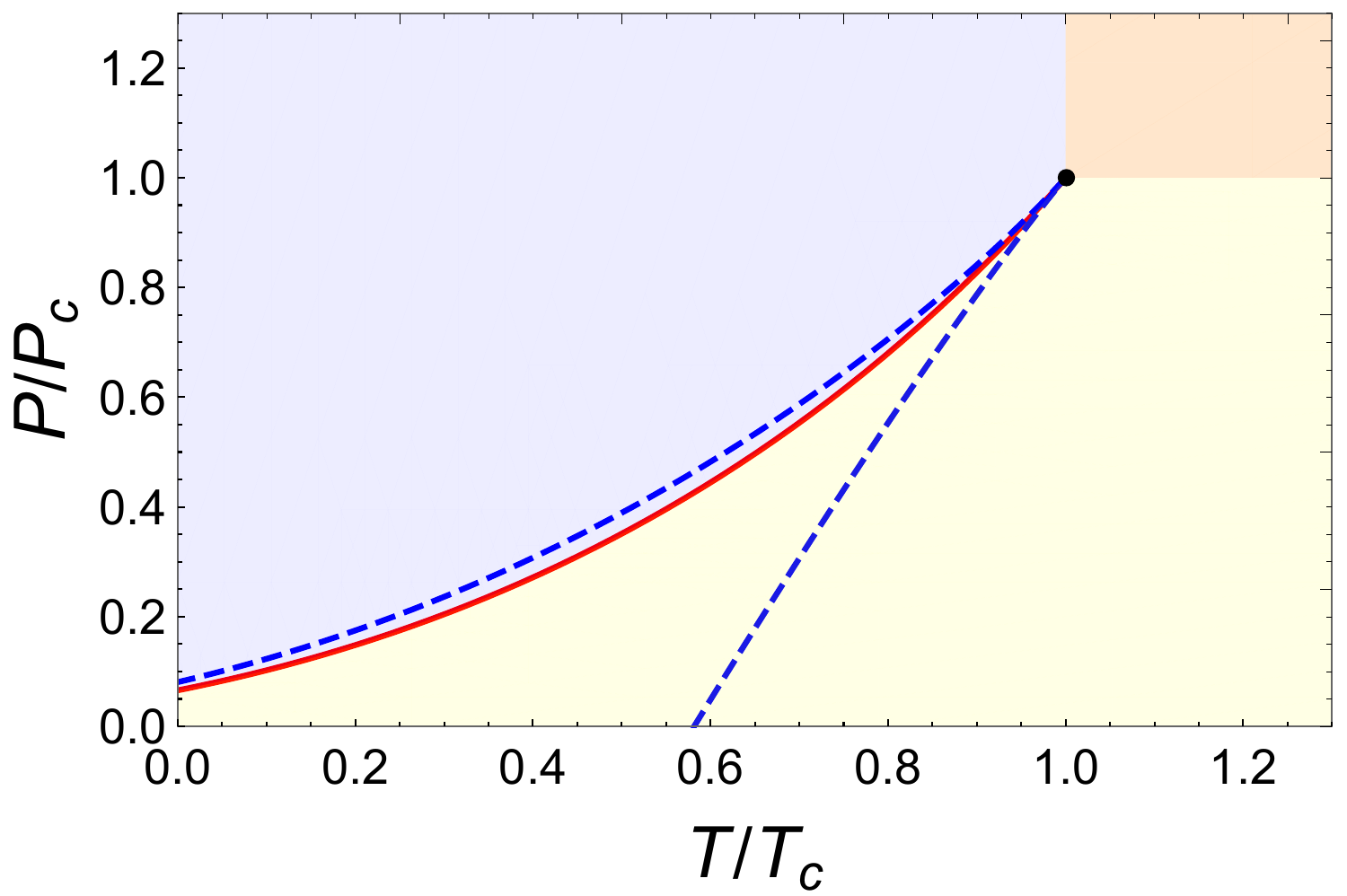}
}
\end{center}
\caption{Phase structure of dRGT black hole in reduced $(\widetilde{P}, \widetilde{T})$ space with $x=0.3$, $0$ and $-0.3$. The solid red and blue dashed curves represent the small\big/large black hole coexistence curve and spinodal curves, respectively.}\label{Fig4}
\end{figure}
The black dote is the critical point $(\widetilde{P}=1,\widetilde{T}=1)$, and the region at the conner marked as orange is the supercritical black hole phase, in which the small black hole and the large black hole are indistinguishable. As shown in the figure, there are two spinodal curves separated by the coexistence curve, and these three curves meet at the critical point. The purple and yellow regions represent the small black holes phase and large black holes phase, respectively.
The shapes of these four figures are similar, but there are some interesting things because of the existence of massive parameters: the starting point of the coexistence curve changes with $x$. The coexistence curve is the separation between the singlet small black hole and the singlet large black hole. For positive $x=0.3$, the coexistence curve starts at $\widetilde{P}=0$ and $\widetilde{T}=3/13$, which implies the absence of the singlet large black hole at sufficiently low temperature. While for negative $x$, the starting point of the coexistence curve is $\widetilde{T}=0$ with finite pressure $\widetilde{P}$, which implies the absence of the singlet small black hole at sufficiently low pressure.
The results restore to RN-AdS when $x=0$ that the coexistence curve starts from the origin of the phase space. Fig.\ref{Fig5} is the plot of the coexistence curve and spinodal curves in the reduced $(\widetilde{T},\widetilde{V})$ phase space with $x=0.3$, $0$ and $-0.3$. The two metastable phases (superheated small black holes and supercooled large black holes) regions and the coexistence phase region are clearly displayed.
\begin{figure}[!h]
\centering
\subfigure[$x=0.3$]{
\includegraphics[width=5cm]{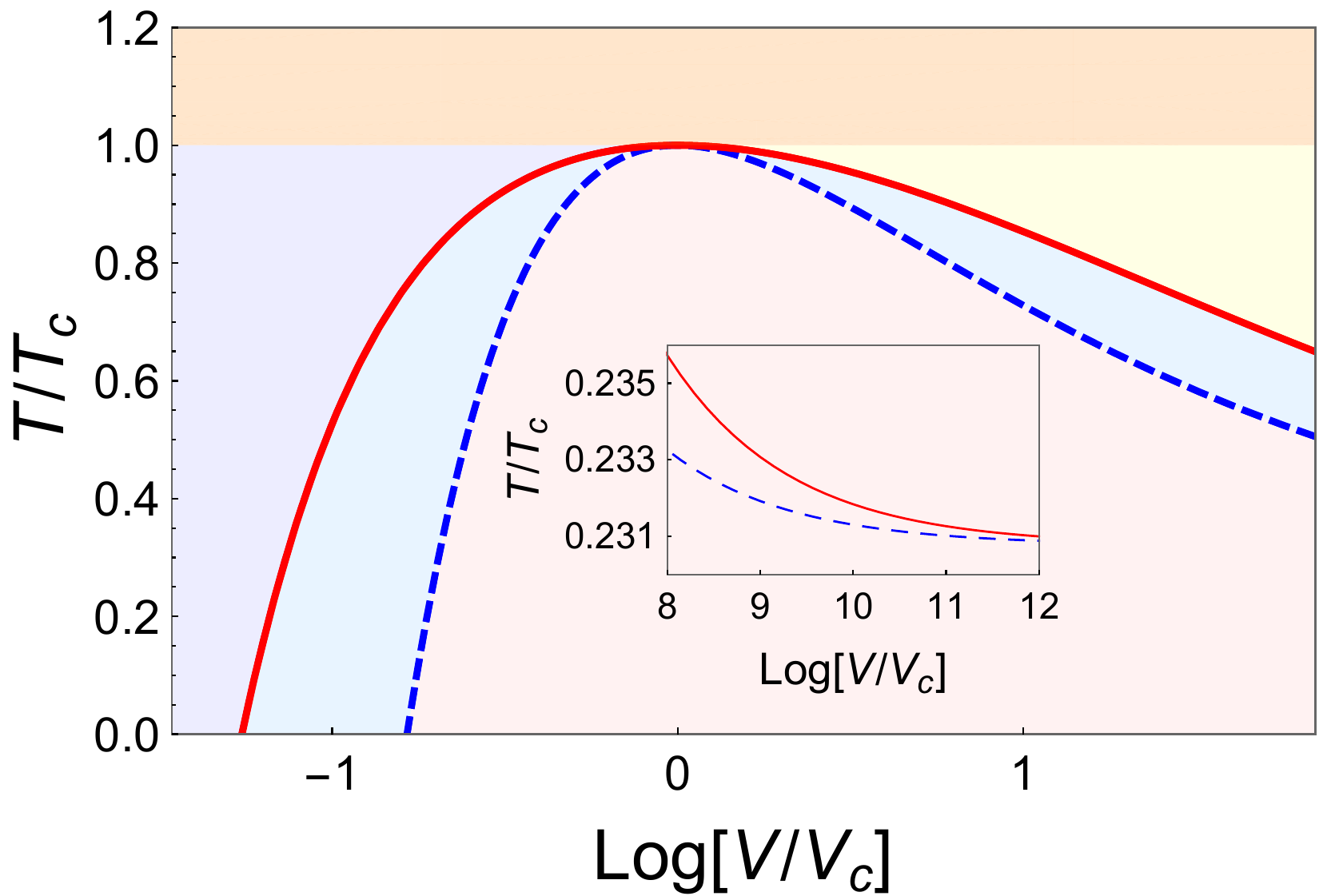}
}
\quad
\subfigure[$x=0$]{
\includegraphics[width=5cm]{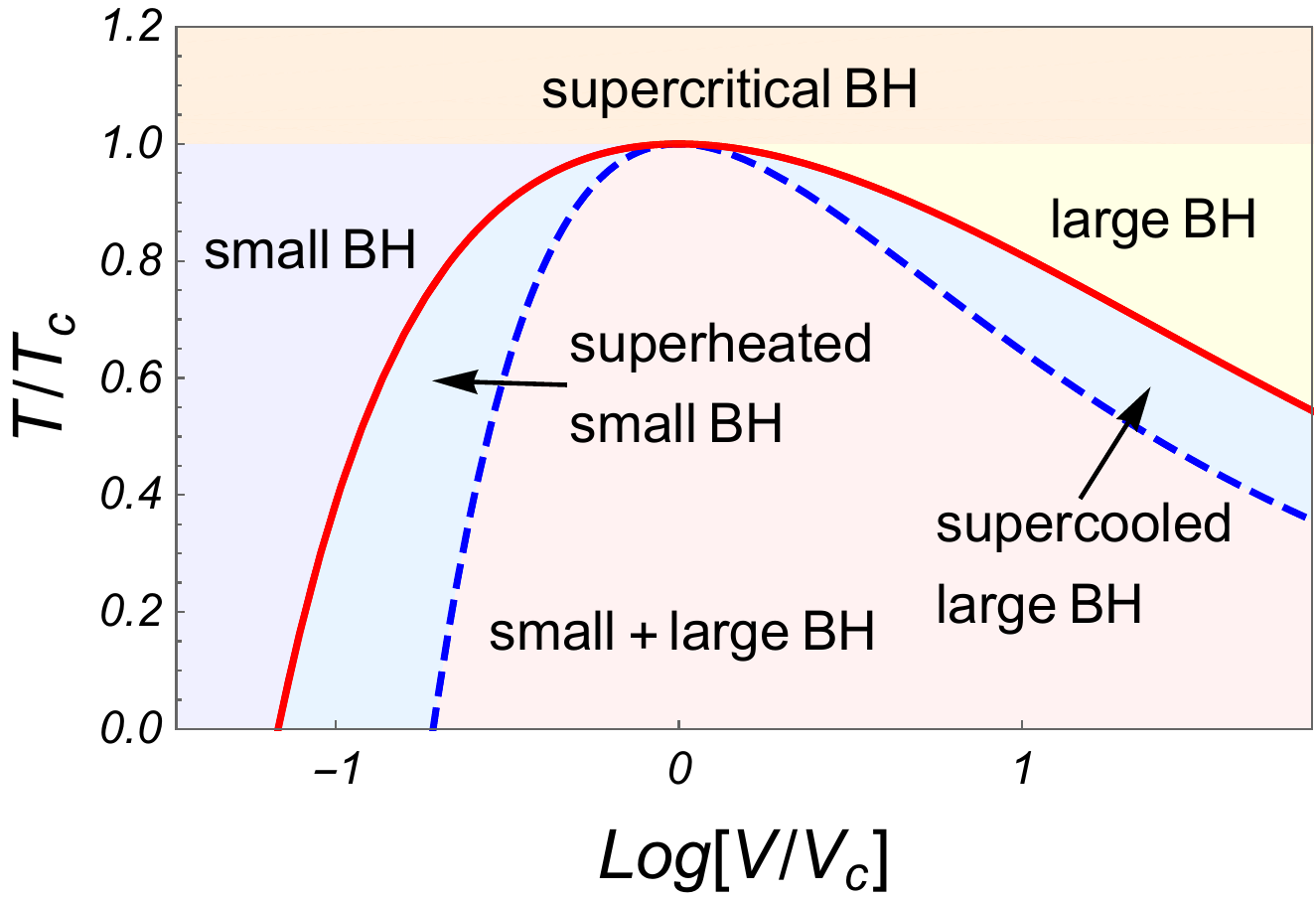}
}
\quad
\subfigure[$x=-0.3$]{
\includegraphics[width=5cm]{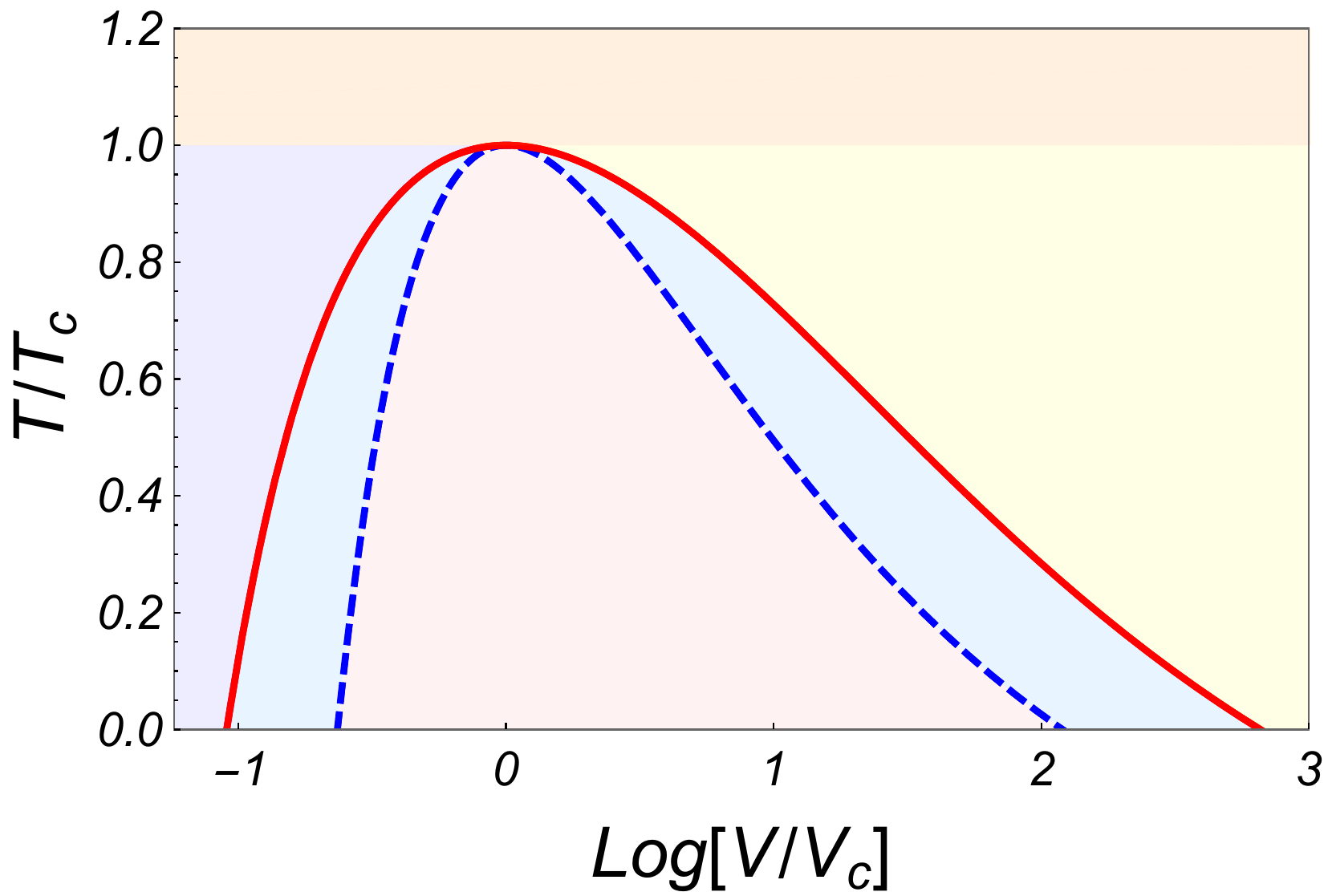}
}
\caption{Phase structure of dRGT black hole in reduced $(\widetilde{T},\widetilde{V})$ space with $x=0.3$, $0$ and $-0.3$. The solid red curve represents the coexistence curve of the small/large black hole, and the blue dashed curve represents the spinodal curve. The regions marked as orange, purple, yellow, green and pink describe the supercritical black hole, small black hole, large black hole, metastable phases (superheated small black hole and supercooled large black hole), small/large black hole coexistence phase, respectively.}\label{Fig5}
\end{figure}
The red solid and blue dashed curves represent the coexistence curve and spinodal curve, respectively. The black holes in different phases are identified with different colors in the figure. We can see that the asymptotic behavior of the coexistence curve at large thermodynamical volume $\widetilde{V}$ is greatly changed by $x$. For $x=0.3$, the coexistence curve asymptotically reaches a finite constant temperature $\widetilde{T}=3/13$ at infinite $\widetilde{V}$, leading to the absence of singlet large black hole at sufficiently low temperature. It implies that there is a critical temperature $\widetilde{T}=3/13$ with positive combined parameter $x=0.3$. Below the critical temperature, there are only two phases (singlet small black hole phase and coexistence phase) with different volume. However, above the critical temperature, three phases (singlet small black hole phase, coexistence phase, and also singlet large black hole phase) of the system with different volumes exist. So that the difference between the small black hole and the large black hole becomes infinite when the temperature $\widetilde{T}\le3/13$, which is consistent with the results indicated by Fig.\ref{Fig3}(b). For $x=0$, the asymptotic value is zero. While for negative $x$, there is an intersection point of the coexistence curve with $\widetilde{T}=0$ horizontal line at finite volume. So that the area of the region under the coexistence curve, becomes small and small as $x$ decreases. We speculate that, as the combined parameter $x$ decreases, a smaller change in thermodynamical volume at a fixed phase transition temperature would cause the transition from singlet small black hole phase to singlet large black hole phase.

\section{\textsf{Ruppenier geometry of the black hole}} \label{IV}
In this section, we concentrate our attention on the thermodynamic geometry of the system and attempt to reveal some information about the microstructure of the black holes. As the situation of RN-AdS black hole, $S$ and $V$ of dRGT black hole are dependent, the heat capacity at constant volume $C_V=T(\partial_T S)_V$ vanishes, which results in a singularity of the thermodynamical metric Eq.(\ref{IM}). Introducing a normalized scalar curvature could be a way for us to explore the thermodynamical properties of the black hole\cite{Wei:2019uqg}, which is defined as
\begin{equation}
    R_N=R C_V=\frac{\left(\partial_V P\right)^2-T^2\left(\partial_{V,T}P\right)^2+2T^2\left(\partial_V P\right)\left(\partial_{V,T,T}P\right)}
                   {2\left(\partial_V P\right)^2}.    \label{sc}
\end{equation}
After a straightforward calculation, the normalized
scalar curvature of dRGT black hole in terms of reduced parameters with $k+c_2>0$ is
\begin{equation}
    R_N=\frac{\left[\frac{c_1 q}{8}\left(\frac{6}{k+c_2}\right)^{\frac{3}{2}}\widetilde{V}
                    +3\widetilde{V}^\frac{2}{3}-1\right]
              \left\{-4\widetilde{T}\widetilde{V}\left[1+\frac{c_1 q}{16}
              \left(\frac{6}{k+c_2}\right)^{\frac{3}{2}}\right]
              +\frac{c_1 q}{8}\left(\frac{6}{k+c_2}\right)^{\frac{3}{2}}\widetilde{V}
              +3\widetilde{V}^\frac{2}{3}-1\right\}}
              {2\left\{-2\widetilde{T}\widetilde{V}\left[1+\frac{c_1 q}{16}
              \left(\frac{6}{k+c_2}\right)^{\frac{3}{2}}\right]
              +\frac{c_1 q}{8}\left(\frac{6}{k+c_2}\right)^{\frac{3}{2}}\widetilde{V}
              +3\widetilde{V}^\frac{2}{3}-1\right\}^2}.    \label{Q11}
\end{equation}
In order to better understand the effect of parameter $x$ on the black holes, we employ Eq.(\ref{Q11}) and set $\widetilde{T}=0.4$ to show the behavior of normalized scalar curvature $R_N$ as the function of $\widetilde{V}$ with $x=0.3$ and $-0.3$ in Fig.\ref{Fig7}.
\begin{figure}[!h]
\centering
\subfigure[$x=0.3$]{
\includegraphics[width=7cm]{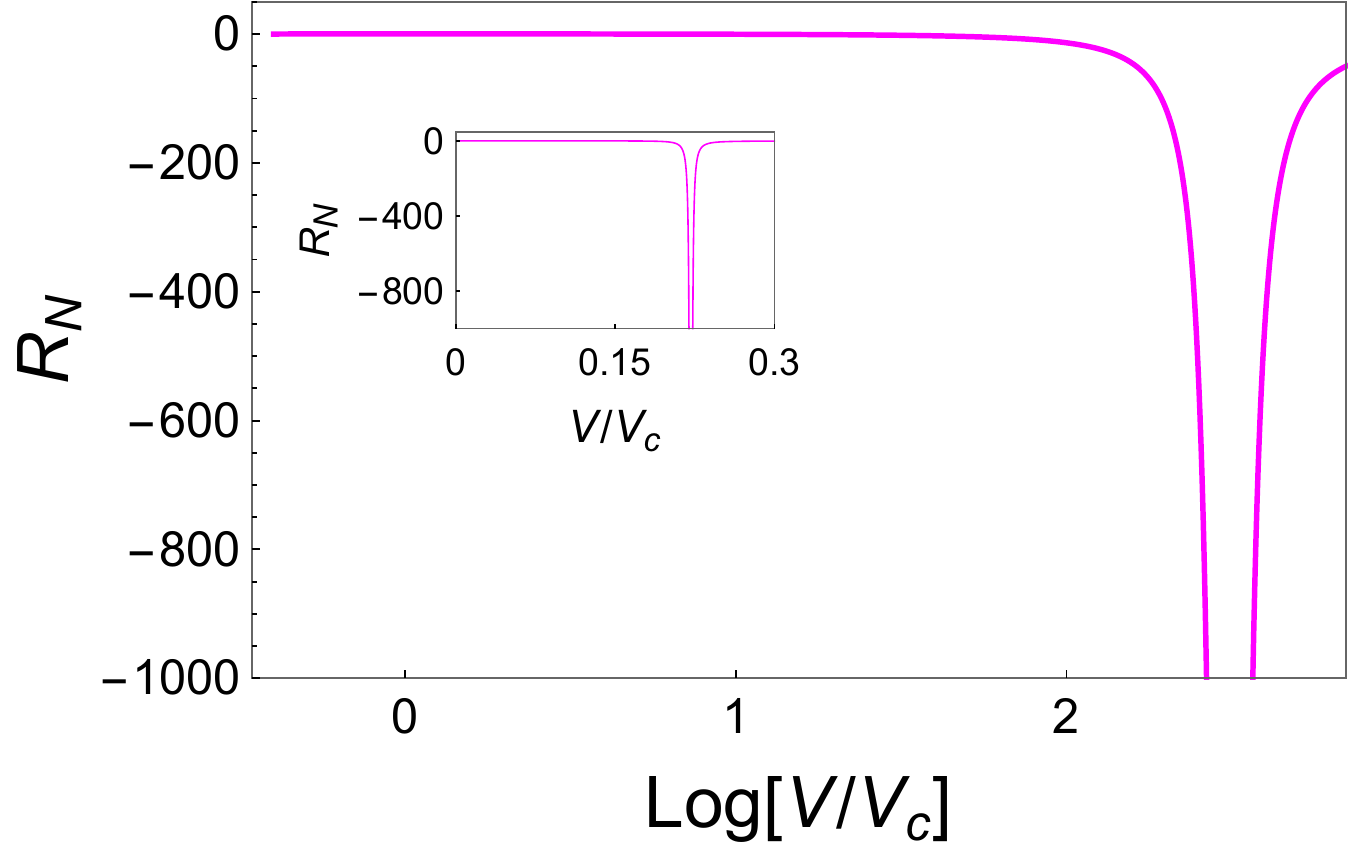}
}
\quad
\subfigure[$x=-0.3$]{
\includegraphics[width=7cm]{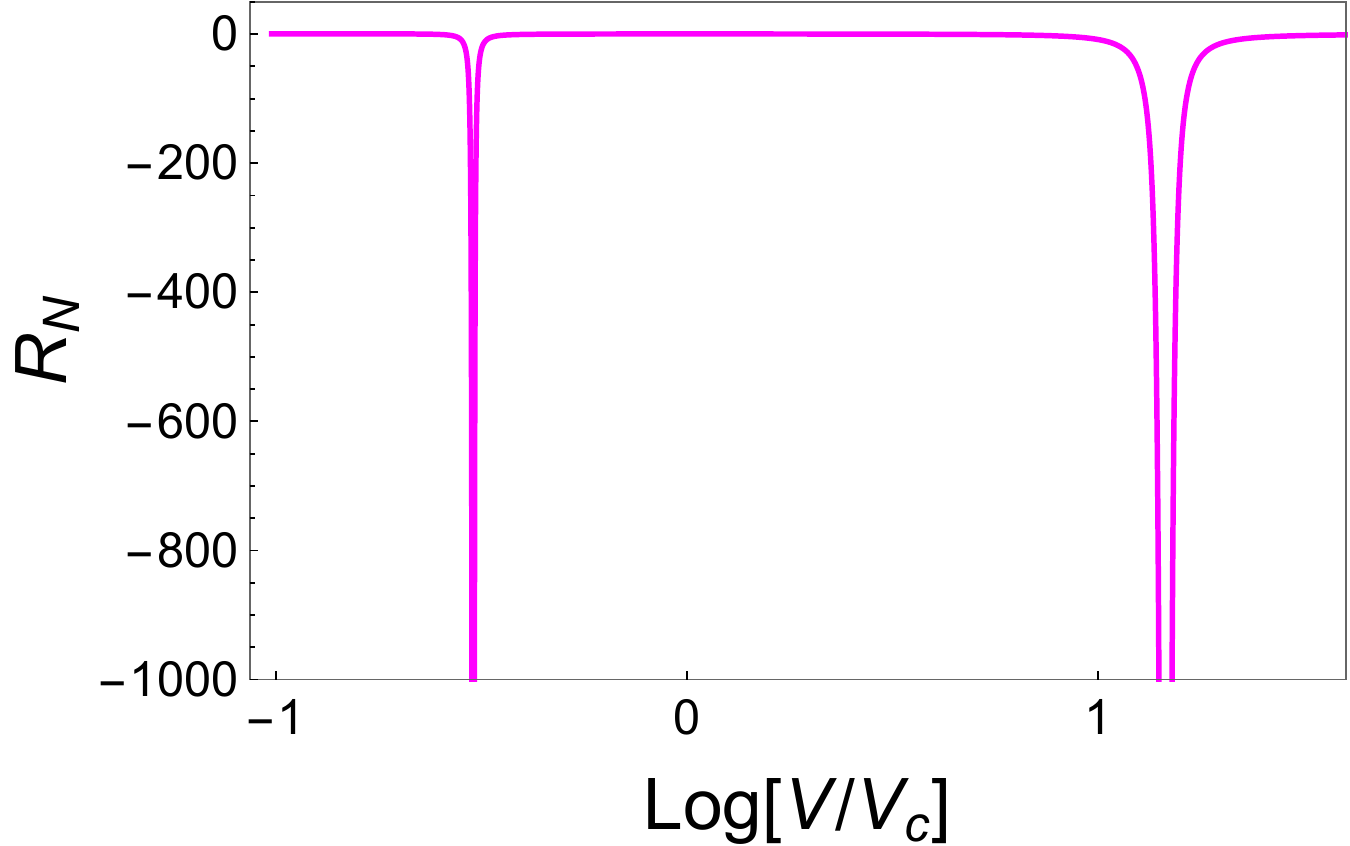}
}
\caption{The normalized scalar curvature of dRGT black hole as function $\widetilde{V}$ with $x=0.3$, $-0.3$, and $\widetilde{T}=0.4$.}\label{Fig7}
\end{figure}
We can see that $R_N$ has two negative divergent points determined by the condition $\partial_V P=0$, which corresponds to the spinodal curve. These two divergent points get close as $x$ decreases. In addition, it is easy to verify from Eq.(\ref{Q11}) that these two divergent points coincide at $\widetilde{V}=1$ when $\widetilde{T}=1$. Fig.\ref{Fig7} shows us that the scalar curvature is mostly negative in the region of parameter space $\widetilde{V}$, but also has the possibility to be positive.

The parameter space $\widetilde{V}$ for a positive curvature with fixed temperature $\widetilde{T}$ and $x$ can be worked out directly by setting $R_N>0$, however, it should be careful to examine where the parameter space is physically allowed. We can check this with sign-changing curves, whose expression are
\begin{align}
    \widetilde{T}_{\rm sc}=&\left[\frac{3\widetilde{V}^\frac{2}{3}-1}{4\widetilde{V}}
                                +\frac{c_1 q}{32}\left(\frac{6}{k+c_2}\right)^{\frac{3}{2}}\right] \Bigg / \left[1+\frac{c_1 q}{16}\left(\frac{6}{k+c_2}\right)^{\frac{3}{2}}\right],    \label{Q16}
\end{align}
and $\widetilde{V}_{\mathrm{sc}}$ determined by the equation
\begin{align}
  \frac{\tilde{c}_1q}{8}\left(\frac{6}{k+\tilde{c}_2}\right)^{\frac{3}{2}}\widetilde{V}_{\mathrm{sc}}
                          +3\widetilde{V}_{\mathrm{sc}}^\frac{2}{3}&-1=0.   \label{Q17}
\end{align}
Before we proceed with plotting the sign-changing curves, we take an analysis of the roots of the above equation.  At first, we inversely solve the equation to the form
\begin{equation}
    x=\frac{1-3\widetilde{V}^\frac{2}{3}}{2\widetilde{V}},    \nonumber
\end{equation}
where the combined parameter $x$ is introduced. Taking the first order derivative of the above equation, we have
\begin{equation}
     \frac{d x}{d \widetilde{V}}=\frac{\widetilde{V}^\frac{2}{3}-1}{2\widetilde{V}^2},    \nonumber
\end{equation}
which indicates that $x$ is monotonically increasing for $\widetilde{V}>1$ and monotonically decreasing for $0<\widetilde{V}<1$. With the asymptotic behaviors
\begin{equation}
    x(\widetilde{V}=0)\rightarrow \infty, \quad x(\widetilde{V}\rightarrow \infty)\rightarrow 0,
\end{equation}
we know that for a fixed positive $x$, the Eq.(\ref{Q17}) admits one solution $\widetilde{V}_{\mathrm{sc}}$, while for a fixed negative $x$, it admits two solutions $\widetilde{V}_{\mathrm{sc}}$; when the parameter $x=0$, we can easily obtain the solution $\widetilde{V}=3^{-\frac{3}{2}}$, which is the result of RN-AdS black hole.
\begin{figure}[!h]
\centering
\subfigure[$x=0.3$]{
\includegraphics[width=5cm]{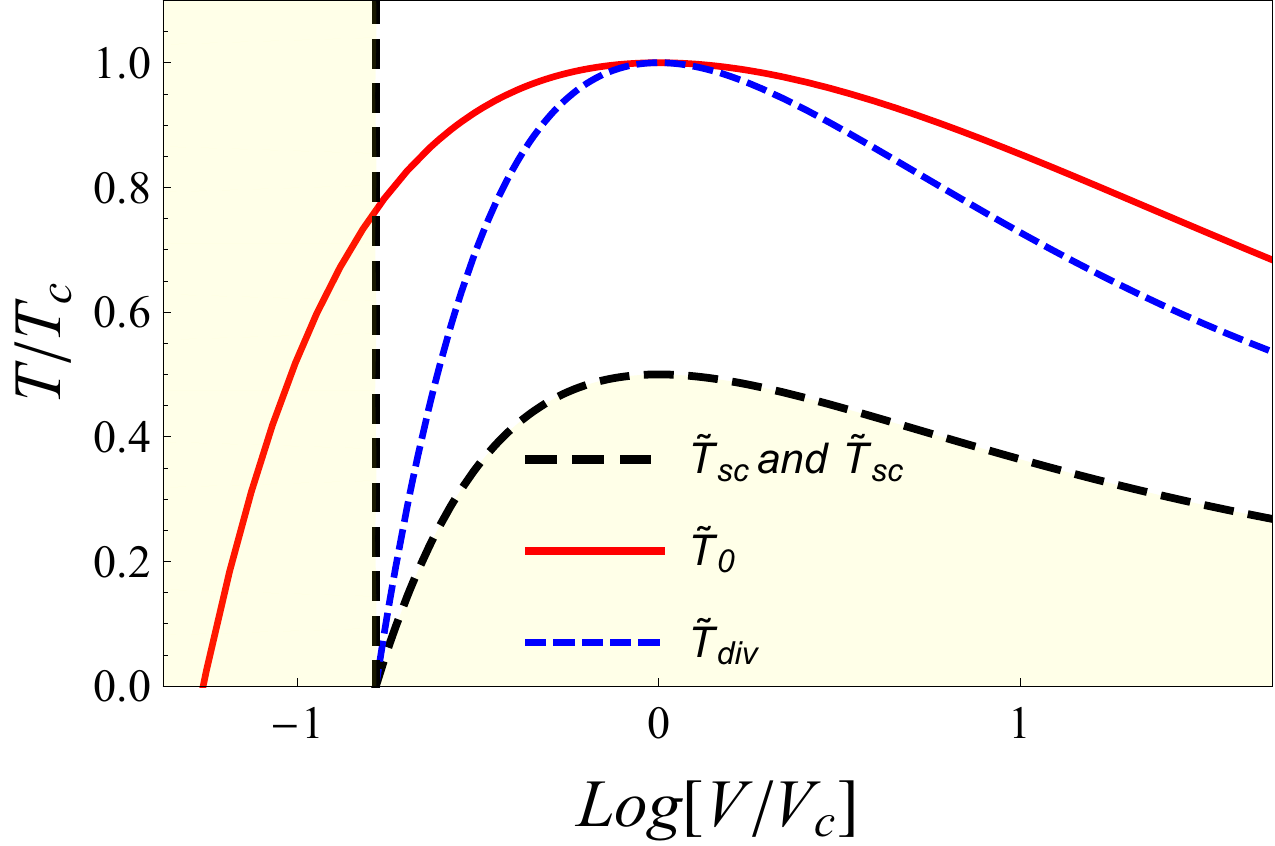}
}
\quad
\subfigure[$x=0$]{
\includegraphics[width=5cm]{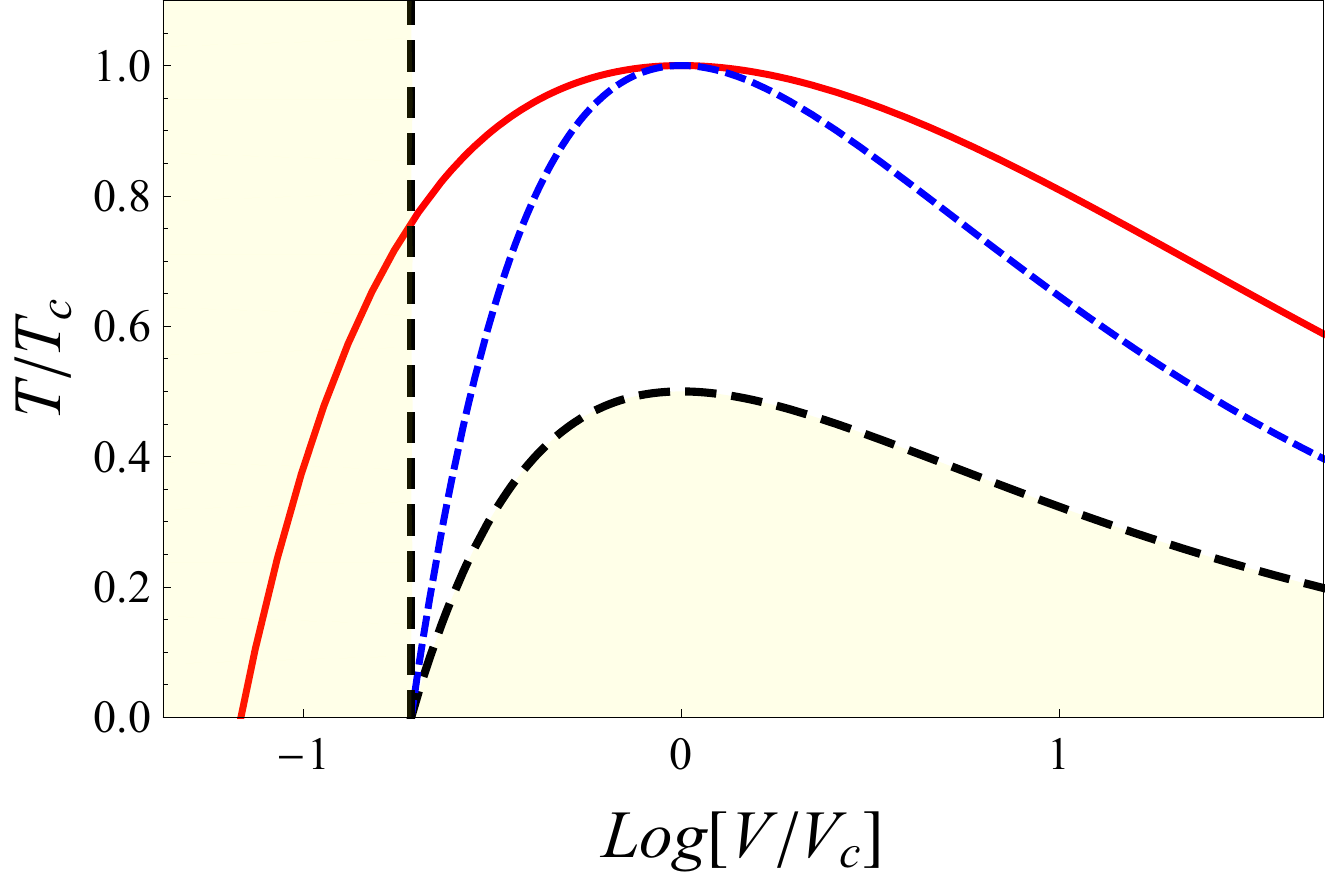}
}
\quad
\subfigure[$x=-0.3$]{
\includegraphics[width=5cm]{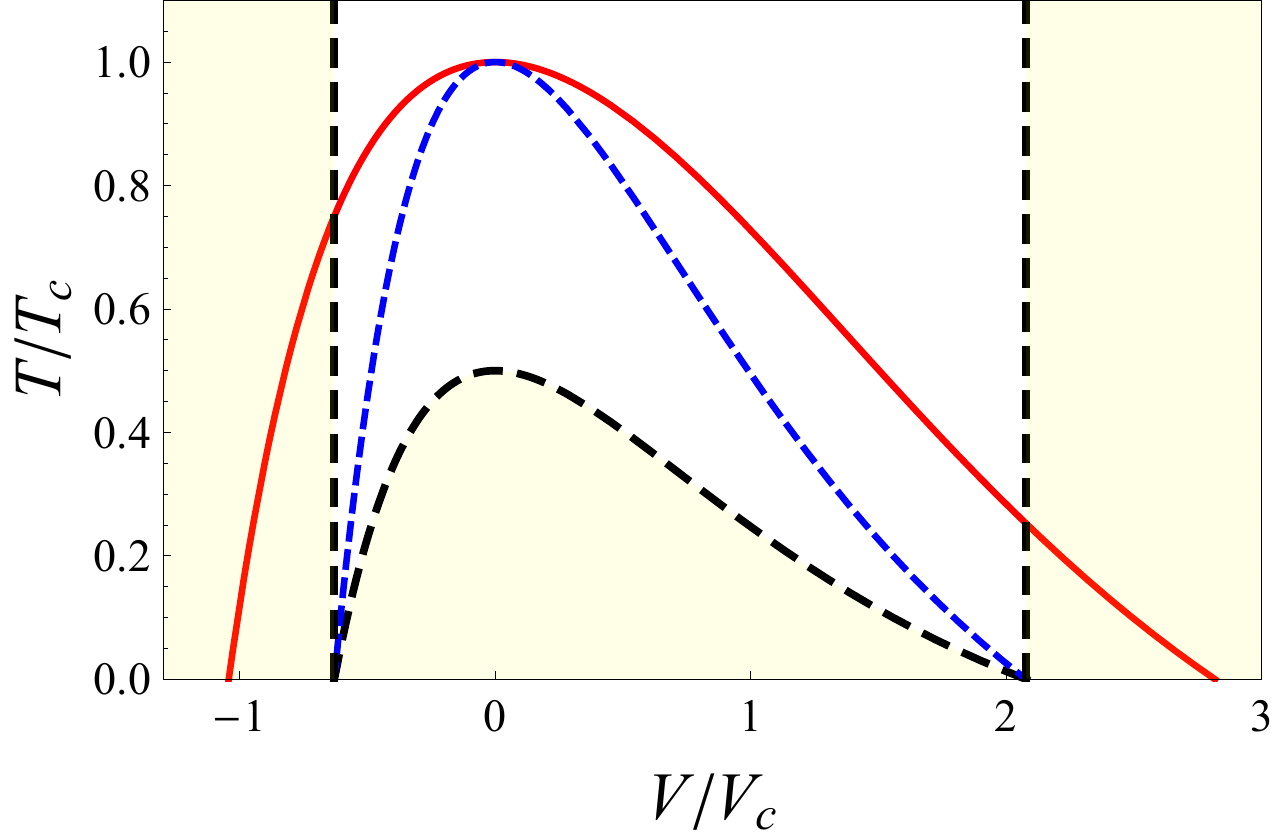}
}
\caption{Characteristic curves of the normalized scalar curvature. The red solid curve is the coexistence curve, and the black and blue dashed curves correspond to the sign-changing curve and divergence curve, respectively. The region painted in yellow represent that $R_N>0$.}\label{Fig9}
\end{figure}
In Fig.\ref{Fig9}, we plot the solid red curve describing the coexistence curve, the blue dashed curve representing the spinodal curve, and dashed black curves representing the sign-changing curves of the normalized scalar curvature $R_N$ with $x=0.3$, $0$ and $-0.3$. The region painted in yellow indicates that the normalized scalar curvature is positive, while the white region is negative. As we have already known from Van der Waals fluid system that the equation of state should be inapplicable in the region  under the coexistence curve. So that the divergent and sign-changing behaviors of scalar curvature in the coexistence phase are excluded from the thermodynamical viewpoint. Only attractive interactions between the molecules of Van der Waals fluid are allowed in the hard-core model\cite{Johnston_2014}. However, as depicted in Fig.\ref{Fig9}, when the combined parameter $x\geq0$, there is a weak repulsive interaction dominating for small black holes which are in the situation that similar to the RN-AdS black hole, and the singlet large black hole phase disappears for sufficient low temperature for positive $x$.
The interesting things happen in the case when $x<0$, which shows us another positive curvature region {\rm III} since the Eq.(\ref{Q17}) has two solutions with $x<0$. According to the discussion in the previous section, compared with RN-AdS black holes, in addition to small black holes, the interaction between underlying `molecules' of large black holes also behave as repulsion. What's more, as $x$ decreases, the area of the region acting as an attraction between the `molecules' also decreases.

The critical behaviors of normalized scalar curvature can be studied to take the series expansion of $R_N$ along the saturated small and large black hole curves near the critical point. With the help of Eq.(\ref{co1}), Eq.(\ref{V1}), Eq.(\ref{V2}), the normalized scalar curvature along the coexistence curve with different value of $x$ are plotted in Fig.\ref{Fig8}.
\begin{figure}[!h]
\centering
\subfigure[$x=0.3$]{
\includegraphics[width=5cm]{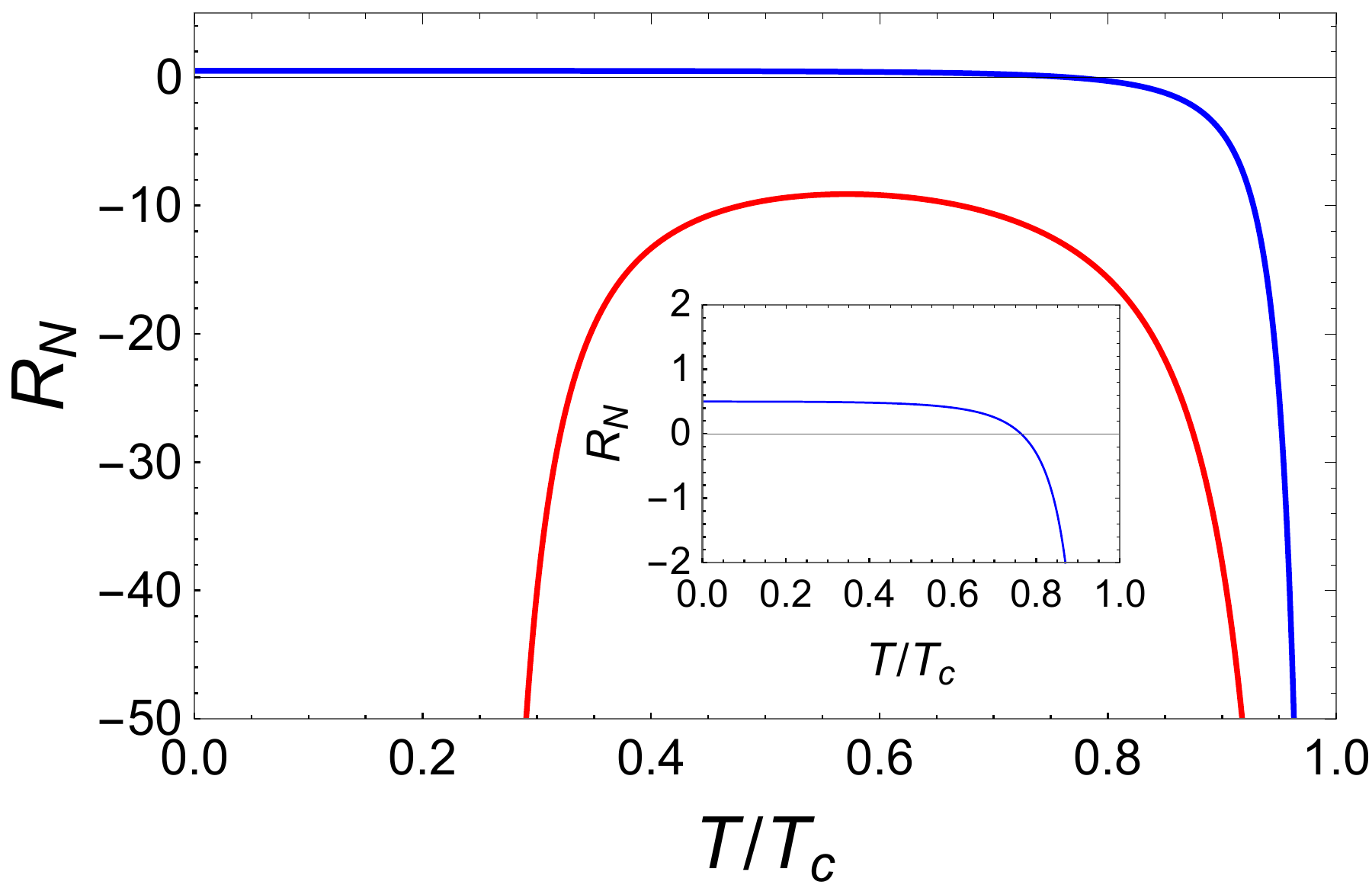}
}
\quad
\subfigure[$x=0$]{
\includegraphics[width=5cm]{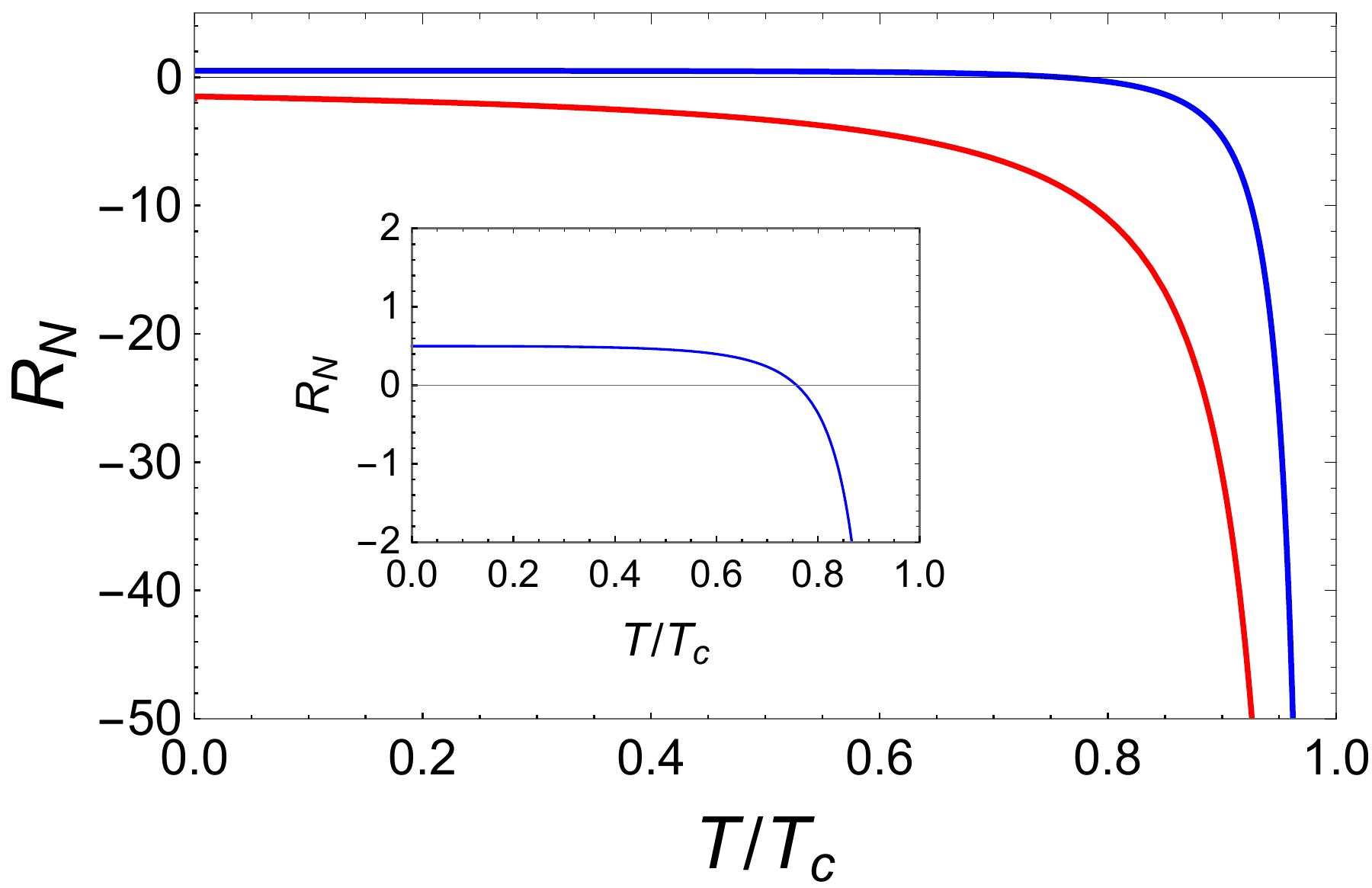}
}
\quad
\subfigure[$x=-0.3$]{
\includegraphics[width=5cm]{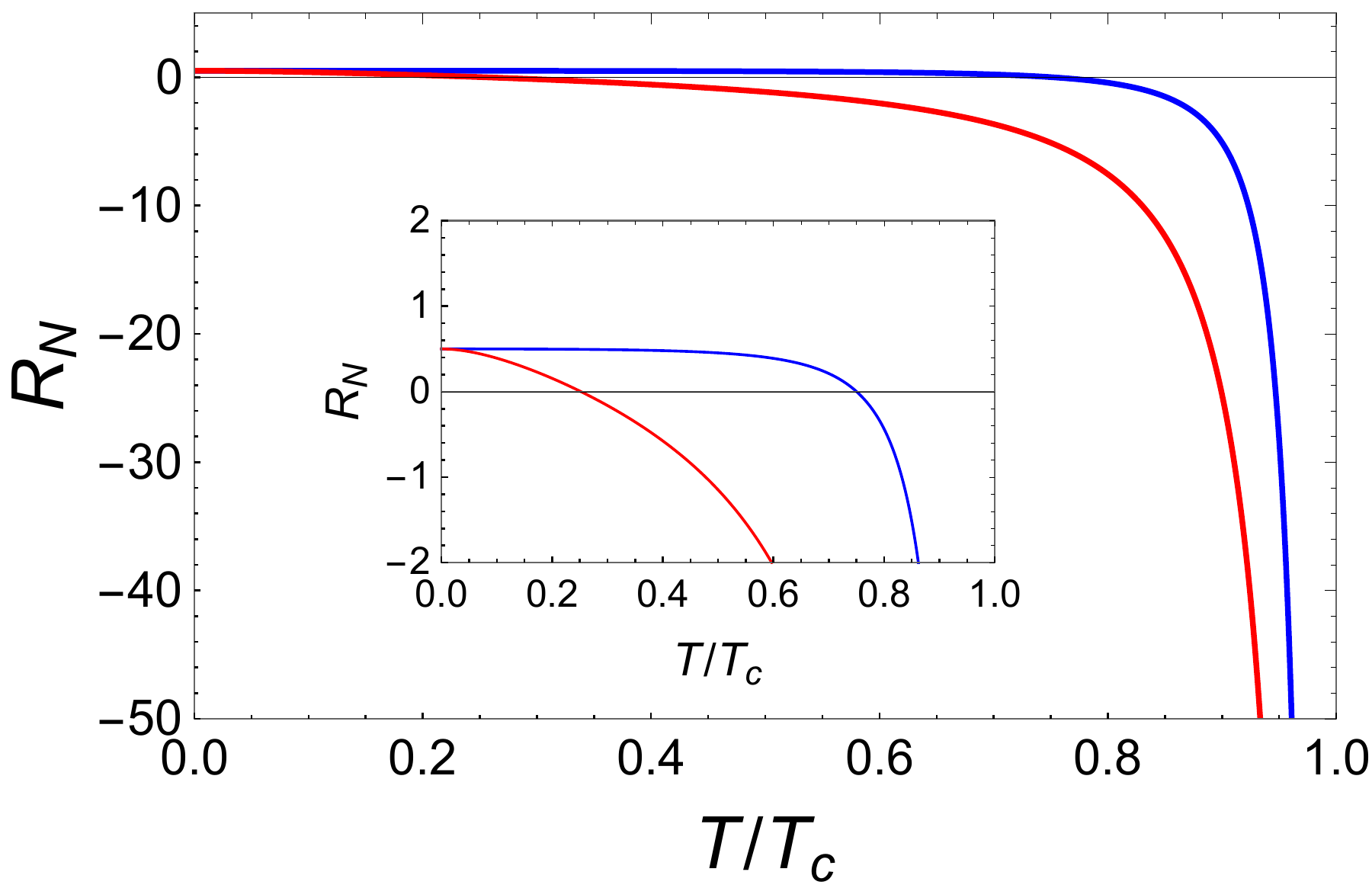}
}
\caption{The behavior of the normalized scalar curvature along the coexistence curvature $R_N$ with $x=0.3$, $0$ and $-0.3$. The solid blue curve represents a saturated small black hole and the red solid curve represents saturated large black hole. The solid red curve in Fig.7(a) tends to be negative infinity at sufficiently low temperature. This behavior is consistent with that of Fig.4(a), where the saturated black hole curve asymptotically approaches to the spinodal curve (where the thermodynamic curvature $R_N$ diverges) at $\widetilde{T}=3/13$.}\label{Fig8}
\end{figure}
The blue (red) curve is $R_N$ along the coexistence curve for saturated small (large) black hole. We can see that the values of the scalar curvature of the large black holes are always smaller than those of the small black holes with fixed temperature, and at the critical point, they all go to negative infinity. Similar to RN-AdS black hole when the combined parameter $x\ge0$, only for the small black hole that $R_N$ can be positive at low temperature. Moreover, for positive combined parameter $x=0.3$, Fig.\ref{Fig8}(a) shows us that $R_N$ along the saturated large black hole curve (solid red line) seems to become negative infinity as the temperature approaching $\widetilde{T} \rightarrow 0$. In fact, there is a truncation of the horizontal axis for the solid red line, the divergent point of $R_N$ for the large black hole (solid red curve) is located at $\widetilde{T}=3/13$. The intriguing behavior of the solid red curve at low temperature in Fig.\ref{Fig8}(a) is consistent with the behavior of the solid red curve in Fig.\ref{Fig5}(a). From the figure of phase structure Fig.\ref{Fig5}(a), we can see that the saturated large black hole curve (solid red line) and spinodal curve (where the thermodynamic curvature $R_N$ diverges) coincide at the truncated temperature $\widetilde{T}=3/13$. The same as Fig.\ref{Fig5}(a), the behaviors of Fig.\ref{Fig8}(a) implies that there is a critical temperature $\widetilde{T}=3/13$ with positive $x=0.3$. With a fixed temperature below the critical value $\widetilde{T}<3/13$, no singlet large black hole phase exists in the system. When the temperature of the system is larger than the critical one, three phases (the singlet small black hole phase, coexistence phase, and also singlet large black hole phase) exist with different volume.

When $x<0$, there would be a vanishing point for $R_N$ of the saturated large black hole, and it moves to a higher temperature as $x$ decreases. Hence the interaction between the `molecules' behaving as repulsion not only dominates in the small black hole at low temperature but also exists in the large black hole with a negative combined parameter $x$. These behaviors are consistent with the analysis from Fig.(\ref{Fig9}).

Finally, we discuss the behavior of curvature towards the critical point along the coexistence curve. For this purpose, we write $R_N$ as a function of $\widetilde{T}$ by using equations Eq.(\ref{co1}), Eq.(\ref{V1}), and Eq.(\ref{V2}), and the series expansions of it at $\widetilde{T}=1$ along the coexistence curves have the following forms
\begin{align}
    R_N(\mathrm{SSBH})=&-\frac{1}{8}(1-\widetilde{T})^{-2}
              +\frac{\sqrt{1+x}}{2 \sqrt{2}} (1-\widetilde{T})^{-\frac{3}{2}}
              + \mathcal{O} (1-\widetilde{T})^{-1},    \nonumber\\
    R_N(\mathrm{SLBH})=&-\frac{1}{8}(1-\widetilde{T})^{-2}
              -\frac{\sqrt{1+x}}{2 \sqrt{2} } (1-\widetilde{T})^{-\frac{3}{2}}
              + \mathcal{O} (1-\widetilde{T})^{-1},    \nonumber
\end{align}
in which SSBH and SLBH represent saturated small black holes and saturated large black holes, respectively. The series expansions of $R_N$ indicate that the critical exponent is $2$. The results can also be checked numerically. Near the critical point, the normalized scalar curvature is assumed to take the form of \cite{Johnston_2014}
\begin{equation}
    \mathrm{ln}|R_N| = -a \, \mathrm{ln} (1-\widetilde{T})+b.   \label{Fit}
\end{equation}
We numerically calculate the curvature along the coexistence curve and list the fitting results in Table \ref{Tab1}. We find that the coefficient $a$ always fluctuates in a small range of around $2$. In the case of the error caused by calculation, we believe that the critical exponent is $2$, which is the same as that of the RN-AdS black hole.
\begin{table}[h]
\begin{center}
\begin{tabular}{@{}lllll@{}}
\toprule
       &$x$=$0.3$ \qquad \quad &  $x$=$0 $  \qquad \quad  & $x$=$-0.3$  \\\midrule
a (SSBH) \qquad \qquad & 1.98695 & 2.03720  & 2.03242  \\
-b (SSBH)\qquad \qquad & 1.91910  & 2.52126 & 2.45151  \\
\hline
a (SLBH)  & 2.07444 & 2.04285 & 2.03545  \\
-b (SLBH) & 2.85544 & 2.51507 & 2.41169  \\ \bottomrule
\end{tabular}
\caption{Coefficients $a$ and $b$ of the fitting formulation along the saturated small black hole (SSBH) and saturated large black hole (SLBH) with $x=0.3$, $0$ and $-0.3$.}
\label{Tab1}
\end{center}
\end{table}

\section{\textsf{Summary and Discussion}}\label{V}

In this paper, we studied the phase transition and thermodynamic geometry of a four-dimensional charged topological black hole in massive gravity. Unlike the RN-AdS black hole that only the spherical topology exhibiting Van der Waals-like phase transition, the so-called small/large black hole phase transition for the charged AdS black holes always exist, no matter the horizon topology is spherical $(k = 1)$, Ricci flat $(k = 0)$ or hyperbolic $(k=-1)$. So that the phase structure of the topological black hole in massive gravity is significantly different from that of the RN-AdS black hole. Specificity, the critical behaviors of topological black hole exhibit provided $k+c_2 > 0$. We find that there does exist a combined parameter in terms of the topology, charge, and massive parameter of the massive black hole, which characterizes the thermodynamic properties of the black hole. The coexistence curve and spinodal curve (the curve separate the metastable phase and unstable phase) are obtained analytically and plotted in the reduced phase space $(\widetilde{P},\widetilde{T})$ and $(\widetilde{T},\widetilde{V})$, respectively. The supercritical phase region, metastable phase (superheated small black holes and supercooled large black holes) regions, and coexistence phase region are clearly displayed. We considered the effect of the crucial combined parameter on the phase structure of the massive black hole. It is found that as the combined parameter $x$ decreasing, the area of the coexistence phase region in the phase space decreases. What's more, we calculated the change of thermodynamical volume of the saturated small black hole and saturated large black hole, the change with a fixed temperature also decreases as the combined parameter $x$ decreasing. It indicates the difference
between small black hole phase and large black hole phase is smaller, thus a smaller change in thermodynamical volume at a fixed phase transition temperature would cause the transition from singlet small black hole phase to singlet large black hole phase. In the case of $x = 0$, the results reduce to that of RN-AdS black hole.

Furthermore, we studied the Ruppeiner thermodynamical geometry of the topological black hole in massive gravity at the phase transition. Since the thermodynamical metric degenerates because of the non-independence of the entropy and thermodynamical volume of the massive black hole, the normalized scalar curvature was introduced to reveal some information about the microstructure of the black hole. As we have shown in Fig.(\ref{Fig7}), the curvature goes to infinity at the critical points, and divergent points (corresponding to the spinodal points) get close as $x$ decreases. One of the most prominent results of the topological massive black hole is that the sign-changing curves of the Ruppeiner curvature are subtle. When the combined parameter is non-negative, the situation is similar to the RN-AdS black hole, i.e., the curvature changes its sign at sufficiently low temperature only for small black holes, that the microstructure interactions could be attractive or repulsive. While for the large black hole, the scalar curvature is always negative, which indicates that among the microstructures for the large black hole there are only attractive interactions. However, when the combined parameter is negative, it is intriguing that both for the small and large black hole there are phase regions admitting positive curvature (See Fig.(\ref{Fig9}) and Fig.(\ref{Fig8})). Also, we found that there is no singlet large black hole at sufficiently low temperature with positive $x$, and a phase transition between the coexistence phase and singlet large black hole occurs at $\widetilde{T}=3/13$ with $x=0.3$. These constitute the distinguishable features of dRGT massive topological black hole from that of RN-AdS black hole as well as the Van der Waals fluid system. It would be interesting to extend our discussion to the higher dimensional dRGT massive gravity that reentrant phase transition, the tricritical point can appear, we leave these issues for future work.

At last, we talk about since in the AdS/CFT correspondence the cosmological constant is normally considered to be a fixed parameter that does not vary, it is natural to ask the interpretation once $\Lambda$ is treated as a thermodynamic variable. It is suggested \cite{Johnson:2014yja,Dolan:2014cja} that varying cosmological constant could be viewed as varying the number of colors $N$ of the CFT. Thus the behaviors of the chemical potential of the boundary field theory can reflect the behaviors of the black holes, such as Hawking-Page transition \cite{Dolan:2014cja} and thermodynamics phase transition of the bulk gravity \cite{Zhang:2014uoa}. Alternatively, interpretation suggested that the number of colors $N$ should be keep fixed, so that one can consider that varying $\Lambda$ corresponds to varying the volume on which the field theory resides \cite{Karch:2015rpa}, the `holographic Smarr relation' and also the $p-\mathcal{V}$ criticality of boundary CFT \cite{Dolan:2016jjc} is studied. In Ref.\cite{Caceres:2015vsa}, the authors argued that varying the cosmological constant provides an alternative notion of changing an energy scale of the dual theory. So that the associated phase structure ($PV$ phase space) of the black hole thermodynamics in the extended phase space is conjectured to be dual to an RG-flow on the space of field theories. They also investigated the behaviors of the entanglement entropy and two-point correlation function of the boundary field theory, which might be good indicators of Van de Waals phase transitions in the bulk black hole thermodynamics. In Ref\cite{Cadoni:2017fnd} the authors also attempted to characterize the shear viscosity to entropy density ratio of dual QFT through Van der Waals-like behaviors of AdS black hole.

However, unlike the Hawking-Page transition for the AdS-black hole, which was later understood as a confinement/deconfinement phase transition in the boundary CFT via AdS/CFT correspondence, the Van der Waals like phase transition (small/large phase transition) behaviors of AdS-black hole, is lack of a suitable holographic interpretation at present. It is worth emphasizing that the details of the field theory interpretation are still, to a large extent, open questions. Furthermore, since the thermodynamic variables of the black hole correspond clearly to the thermodynamic variables of the dual CFT, we believe that the Ruppeiner thermodynamics geometry of AdS-black hole of the Einstein gravity theory or the modified gravity theory may also have its counterpart of the dual QFT, which is an intriguing and meaningful topic worthing to investigate.

\section*{Acknowledgment}
Bin Wu would like to thank Wei Xu and De-cheng Zou for their helpful discussions.
The financial supports from the National
Natural Science Foundation of China (Grant Nos. 11947208, 12047502),
China Postdoctoral Science Foundation (Grant Nos. 2017M623219,  2020M673460),
Major Basic Research Program of Natural Science of Shaanxi
Province (Grant No. 2019JQ-081), Shaanxi Postdoctoral Science and Double First-Class
University Construction Project of Northwest University are
gratefully acknowledged. The authors thank the anonymous referees
for the constructive suggestions that improve this work greatly.

\section*{Note Added}
Some of the results presented in this paper have also been independently obtained in Ref.\cite{Yerra:2020oph}, which appeared in parallel on arXiv.org.

\end{spacing}

\providecommand{\href}[2]{#2}\begingroup\raggedright\endgroup

\end{document}